\documentclass[12pt,preprint]{aastex}
\def\msun{{\rm ~M}_{\odot}}

\usepackage{lscape}

\usepackage{amsmath,amssymb}  

\usepackage[usenames,dvips]{color}

\begin{document}

\title{X-ray and Gamma-ray Emissions from    Different Evolutionary 
Stage of Rotation  Powered Millisecond Pulsars}
\author{J. Takata\altaffilmark{1}
\email{takata@hku.hk}
K. S. Cheng\altaffilmark{1}
\email{hrspksc@hkucc.hku.hk}
\and
Ronald E. Taam\altaffilmark{2,3}
\email{r-taam@northwestern.edu}}
\altaffiltext{1}{Department of Physics, University of Hong Kong, 
Pokfulam Road, Hong Kong}
\altaffiltext{2}{Department of Physics and Astronomy, Northwestern University,
2131 Tech Drive, Evanston, IL 60208}
\altaffiltext{3}{Academia Sinica Institute of Astronomy and Astrophysics - TIARA,
P.O. Box 23-141, Taipei, 10617 Taiwan}

\begin{abstract}
The $Fermi$-LAT has revealed that rotation powered millisecond pulsars (MSPs) are a major contributor to the Galactic $\gamma$-ray source population. Such pulsars may also be important in modeling the quiescent state of several low mass X-ray binaries (LMXBs), where optical observations of the companion star suggest the possible existence of rotation powered MSPs. To understand the observational properties of the different evolutionary stages of MSPs, the X-ray and $\gamma$-ray emission associated with the outer gap model is investigated.  For rotation powered MSPs, the size of the outer gap and the properties of the high-energy emission are controlled by either the photon-photon pair-creation  process or magnetic pair-creation process near the surface. For these pulsars, we find that the outer gap model controlled by the magnetic pair-creation process is preferable in explaining the possible correlations between the $\gamma$-ray luminosity or non-thermal X-ray luminosity versus the spin down power. For the accreting MSPs in quiescent LMXBs, the thermal X-ray emission at the neutron star surface resulting from deep crustal heating can control the conditions in the outer gap.  We argue that the optical modulation observed in the quiescent state of several LMXBs originates from the irradiation of the donor star by $\gamma$-rays from the outer gap.  In these systems, the irradiation luminosity required for the optical 
modulation  of the source  such as SAX~J1808.4-3658 can be achieved for a neutron star of high mass. Finally, we discuss the high-energy emission associated with an intra-binary shock in black widow systems, e.g. PSR B1957+20.
\end{abstract}


\section{Introduction}
The $Fermi$ Large Area Telescope ($Fermi$-LAT) has discovered over  80 $\gamma$-ray pulsars, and 
revealed that the $\gamma$-ray pulsars are a major class of Galactic $\gamma$-ray sources. Of 
these, $Fermi$-LAT first detected pulsed $\gamma$-ray emission from 11 millisecond pulsars, 
hereafter denoted as MSPs (Abdo et al. 2010a,b,c, 2009a,b; Saz Parkinson et al. 2010; Guillemot et al. 2011).  The 
$\gamma$-ray emission from pulsars has a long history, having been discussed in the context of 
a polar cap accelerator (Ruderman \& Sutherland  1975; Daugherty \& Harding 1982, 1996), a slot gap  
(Arons 1983; Muslimov \& Harding 2004; Harding, Usov \& Muslimov 2005; Harding et al. 2008; 
Harding \& Muslimov 2011) and an outer gap accelerator (Cheng, Ho \& Ruderman 1986a,b; Hirotani 
2008; Takata, Wang \& Cheng 2010b).  The cut-off features of the $\gamma$-ray spectra of the 
Crab and the Vela pulsars as measured by $Fermi$ imply that the $\gamma$-ray emission site of 
canonical pulsars is located in the outer magnetosphere. This is in contrast to a site near the 
polar cap region which produces a cut-off feature steeper than observed (Aliu et al. 2009; Abdo et 
al. 2009c, 2010d). However, Venter, Harding \& Guillemot (2009)  found that the observed 
pulse profiles of several MSPs detected by the $Fermi$ cannot be explained by the outer gap and/or 
the slot gap models. Hence, they proposed a pair-starved polar cap model in which particles are 
continuously accelerated to high altitude because multiplicity of the pairs is insufficient to screen 
the electric field. Therefore, the origin of the $\gamma$-ray emission from MSPs remains to be clarified. 

\begin{figure}
\epsscale{.80}
\plotone{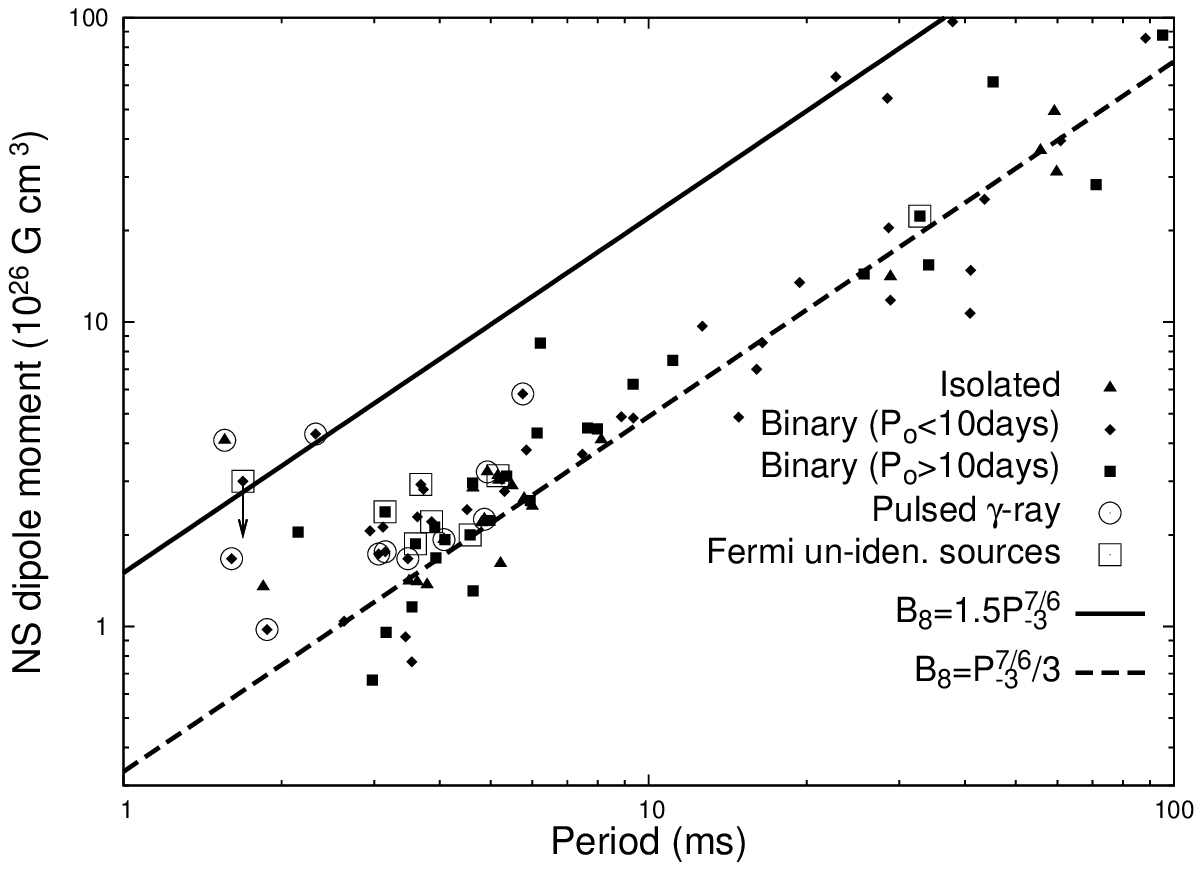}
\caption{Spin period, P, vs. $\mu$ for the rotation powered MSPs. The diamond with the vertical arrow 
represents the recently activated MSP, PSR~J1023+0038 (Archibald et al. 2009). The open circles represent 
the $\gamma$-ray MSPs detected by the $Fermi$, and the open boxes represent the radio MSPs, whose  positions are  coincident  with 
the $Fermi$ unidentified sources. 
The data are taken from ATNF catalog 
(Manchester et al. 2005). The solid and the dashed lines correspond to the relation $\mu_{26}=1.5P_{-3}^{7/6}$ 
and $\mu_{26}=P_{-3}^{7/6}/3$, respectively (see text for details).}
\label{pb}
\end{figure}

We present the spin period vs. the dipole moment for the rotation powered MSPs in Figure~\ref{pb}, 
in which the $\gamma$-ray emitting MSPs are marked with open circles.  
 The spin period in the accretion stage may be related to the equilibrium spin period, which is obtained by equating the co-rotation radius $r_{co}=(GMP^2/4\pi^2)^{1/3}$ to the Alfven 
radius $r_M=(\mu^4/2GM\dot{M})^{1/7}$, where $M$ is the neutron star mass and $\dot{M}$ is the 
accretion rate. The equilibrium spin period implies $\mu\propto P^{7/6}$ (Alpar et al. 1982).  In Figure~\ref{pb}, we 
illustrate the relation $\mu_{26}=\kappa P^{7/6}_{-3}$, where $P_{-3}=(P/1~\mathrm{ms})$,  $\mu_{26}=
(\mu/10^{26}~\mathrm{G~cm^3})$.  The normalization factor  is chosen  as  $\kappa=1.5$ (solid 
line), which  gives an upper limit to the magnetic field for the recently activated radio 
MSP PSR~J1023+0038 (see below),  and $\kappa=1/3$ (dash line), which reproduces the typical $P-\mu$ 
relation for  the Fermi-LAT MSPs and the radio MSPs.  In section~\ref{accretion}, we apply the
above relations to estimate the magnetic moment of accreting MSPs. 

In Figure~\ref{pb}, the new radio MSPs, whose locations are  coincident with the $Fermi$ unidentified 
sources are marked with boxes.  In fact, it has been pointed out that over 20 new radio MSPs have been 
discovered as $Fermi$ identified sources (e.g. Keith et al. 2011).  Since the detection of the spin 
period of MSPs (in particular in binary systems) via a blind search is very difficult, it is likely 
that $Fermi$ has missed the identification of many MSPs.  Thus, some of them may be associated with 
the $Fermi$ unidentified sources.  For example, Takata, Wang and Cheng (2011a,b,c) performed 
population studies of $\gamma$-ray pulsars and argued on statistical grounds that the $\gamma$-ray 
emission from over 100 MSPs are missed, possibly contributing to the $Fermi$ unidentified sources.

PSR~J1023+0038 is known to be the first and only rotation powered radio MSP in a quiescent 
low-mass X-ray binary (LMXB) FIRST J102347.67+003841.2.  The possibility of $\gamma$-ray 
emission from the newly born MSP PSR~J1023+0038 has been pointed out by Tam et al. (2010). 
This system showed clear evidence of an accretion disk in 2001 and its possible absence in 
2002 (Wang et al. 2009), perhaps indicating that the pulsar became activated.  To facilitate 
the transition from an accreting MSP X-ray pulsars (AMP) to a rotation powered MSP, the accretion 
of matter onto the NS must decrease rapidly.  Campana et al. (1998) argued that the ``propeller'' 
effect, in which the Alfven radius exceeds the co-rotation radius, operates during the quiescent 
state of an X-ray transient LMXB phase (Romanova et al. 2009) to eject matter from the system. 
Takata, Cheng \& Taam (2010a) suggest that the activation of the rotation powered pulsar phase 
in a short period LMXB is likely to occur during the quiescent state and that the $\gamma$-ray 
emission produced in the outer gap accelerator in the pulsar magnetosphere irradiates the 
surrounding disk, thereby, further enhancing this ejection of disk material. 

The activation of rotation powered pulsars has been hypothesized based on the orbital 
modulations in the optical emission observed in the quiescent state of LMXBs, i.e., First 
102347.68+003841.2 (Thorstensen \& Armstrong 2005), SAX J1808.4-3658 (Burderi et al. 2003; Deloye 
et al. 2008), XTE~1814-338 (D'Avanzo et al. 2009), and IGR~J00291+5934 (Jonker et al. 2008). 
In these systems, the amplitude of the optical modulation can not be explained by irradiation 
associated with the X-ray emission from the disk or from the neutron star (NS) surface due to 
insufficient luminosity.  To provide an explanation of the orbital modulation of the 
optical emission, pulsar wind models have been suggested in which heating of the donor is 
due to the effect of a relativistic pulsar wind (Burderi et al. 2003).
 However there are some unresolved issues on this picture; for example,   
(1) heating process of the stellar matter  by the pulsar magnetic field  if the wind energy near the companion star 
 is dominated by the magnetic energy  
  and (2) conversion mechanism from  electromagnetic
 energy into the particle energy within $r\sim 10^{10-11}$~cm from the pulsar  if the wind energy is dominated 
 by the particle energy. In this paper,  because of the large theoretical uncertainties on the pulsar wind model,    
we will not pursue this picture in further detail.  Alternatively,  we will discuss
 the $\gamma$-ray irradiation from the outer gap as a possible heating process of the companion star
  (section~\ref{accretion}).

In the X-ray band, the emission from rotation powered MSPs can be composed of thermal emission 
from a heated polar cap region plus (for some pulsars) non-thermal emission of magnetospheric 
origin.  Although the number of the non-thermal X-ray emitting MSPs have increased due to the 
improved sensitivity of recent X-ray instruments (e.g. Zavlin 2007), the origin of pulsed 
non-thermal X-ray emission from MSPs is not well understood.  
Since the population of the $\gamma$-ray MSPs has increased due to the 
detections by the  $Fermi$-LAT, a study of the emission process in the X-ray and  $\gamma$-ray 
bands is necessary for discriminating between various  non-thermal emission processes in 
the magnetosphere. 

Observationally, X-ray emission from the NS surface has been detected during the quiescent state of 
several LMXBs (see Heinke et al. 2009 and reference therein). This emission coupled with estimates for the time averaged mass accretion rate 
has been utilized to probe the equation of state of NS matter (e.g. Yakovlev, Levenfish \& Haensel 
2003; Campana et al. 2008). For some systems, the thermal X-ray emission may be very weak in the 
quiescent state (e.g., SAX~J1808.4-3658) and the X-ray emission may be well fit by a power law 
spectrum (Heinke et al. 2009). The finding of non-thermal X-ray emission from SAX~J1808.4-3658 may 
provide possible evidence for the activation of the rotation powered activity in the quiescent state, 
although the origin of this power law component is unclear at present, as it may be due, for example, 
to the emission from the  magnetosphere or intra-binary shock.

The X-ray emission from an intra-binary shock in the  MSP and low mass star (hereafter LMS) system 
has been suggested to explain the observed unresolved non-thermal  X-ray emissions from PSR~B1957/LMS 
binary system (Arons \& Tavani 1993; Stappers et al. 2003; Huang \& Becker 2007), in which the LMS 
eclipses the radio emission from the MSP.  Moreover,  similar binary (so called ``black widow'') 
systems have been discovered at the positions of the $Fermi$ unidentified sources 
(Roberts et al. 2011), which raises 
questions regarding the origin of the high-energy  emissions from MSP/LMS system.

As we have described above, observational evidence for non-thermal X-ray and $\gamma$-ray 
emission from the different evolutionary stages of MSPs has been accumulating. In this paper, 
we discuss the $\gamma$-ray and X-ray emissions from isolated rotation powered MSPs and 
those in binary systems. Specifically, we model the $\gamma$-ray emission from the outer gap 
of the MSPs, and explore (1) the origin of the $\gamma$-ray emission from rotation powered MSPs 
detected by $Fermi$-LAT and (2) the possibility that irradiation by the $\gamma$-rays 
from the outer gap can explain the optical modulation of quiescent LMXBs.  We also discuss 
the non-thermal X-ray emission associated with the outer gap activities and 
the high-energy emission from an intra-binary shock.  In section~\ref{model}, 
we review the $\gamma$-ray emission from the outer gap accelerator in a pulsar magnetosphere 
and suggest that the thermal X-ray emission from the polar cap region can arise from the 
heating due to incoming particles, which were accelerated in the outer gap.  In 
section~\ref{rotation}, we apply the model to the rotation powered MSPs and compare the 
predicted $\gamma$-ray luminosity with the $Fermi$ observations.  In section~\ref{accretion}, 
the various processes determining the NS surface temperature which depend on the 
NS model of the accreting millisecond pulsars (AMPs) in quiescent LMXBs are discussed.  In 
addition, we estimate the $\gamma$-ray 
luminosity from the outer gap accelerator.  In section~\ref{nonx}, the non-thermal X-ray 
emission is discussed within the context of the outer gap model, and we predict the relation 
between the non-thermal X-ray luminosity and the spin down power of rotation powered MSPs. In 
addition, the non-thermal X-ray luminosity from the AMP in the quiescent LMXB is discussed as 
a function of the time averaged accretion rate. 
In section~\ref{intra}, we discuss  high-energy  emission from an intra-binary shock in MSP/LMS 
binary, and describe the difference in the  properties between the intra-binary 
shock emission  and the magnetospheric emission.  Finally, a brief summary is presented in 
the last section. 

\section{X-ray and $\gamma$-ray radiation processes}
\label{model}
\subsection{$\gamma$-ray emissions from the outer gap}
In the outer gap accelerator model for rotation powered pulsars, electrons and/or positrons 
can be accelerated by the electric field along the magnetic field lines in the region where the 
local charge density deviates from the Goldreich-Julian charge density. The typical strength of the 
accelerating field in the gap is expressed as (Zhang \& Cheng 1997)
\begin{equation}
E_{||}\sim \frac{f^2V_{a}}{s}\sim 9\times 10^5f^2 \left(\frac{P}{10^{-3}~\mathrm{s}}\right)^{-3}
\left(\frac{\mu}{10^{26}~\mathrm{G~cm^3}}\right)\left(\frac{s}{R_{lc}}\right)^{-1}\mathrm{(c.g.s.)},
\label{electric}
\end{equation}
where $V_a=\mu /R_{lc}^2$ is the electrical potential drop across the polar cap, $P$ is the spin period, $\mu$ is the dipole moment of the NS, $s$ is the curvature radius of the magnetic 
field line and $R_{lc}=cP/2\pi$ is the light cylinder radius. In addition, $f$ is the fractional 
gap thickness, which is defined as the ratio of the gap thickness at the light cylinder 
to the light cylinder radius $R_{lc}$. From the energy balance between the dipole radiation and the spin down of the pulsar, the dipole 
moment of the NS is estimated from 
\begin{equation}
\mu \sim \sqrt{3\pi}I^{1/2}R_{lc}^{3/2}P^{-1}\dot{P}^{1/2}\sim
3\times 10^{26}\sqrt{P_{-3}\dot{P}_{-19}}~\mathrm{G~\mathrm{cm^3}}
\end{equation}
where $I$ is NS moment of inertia assumed to be $I=10^{45}~\mathrm{g~cm^2}$.  In addition, 
$\dot{P}_{-19}$ is the time derivative of the spin period in units of $10^{-19}~\mathrm{s/s}$.

By assuming force balance between the acceleration and radiation reaction of the curvature radiation 
process, the typical Lorentz factor of the particles in the gap is 
\begin{equation}
\Gamma=\left(\frac{3s^2}{2e}E_{||}\right)^{1/4}\sim 2\times 10^7 f^{1/2}
P^{-1/4}_{-3}\mu_{26}^{1/4}s_*^{1/4},
\label{gamma}
\end{equation}
where $s_*=(s/R_{lc})$.  The typical energy of the curvature 
photons  is estimated to be  
\begin{equation}
E_{c}=\frac{3hc\Gamma^3}{4\pi s}\sim 25f^{3/2}P_{-3}^{-7/4}\mu_{26}^{3/4}s_{*}^{-1/4}~\mathrm{GeV}.
\label{lcut}
\end{equation}
The $\gamma$-ray luminosity from the outer gap is typically 
\begin{equation}
L_{\gamma}\sim f^3L_{sd}=3.8\times 10^{35}f^3P_{-3}^{-4}\mu_{26}^2~\mathrm{erg~s^{-1}},
\label{lgamma}
\end{equation}
where  we used the spin down power $L_{sd}=2(2\pi)^4\mu^2/3P^4c^3$. 

 We note  that 
the electron and positron pairs are mainly created around 
the inner boundary of the outer gap (Cheng, Ruderman \& Zhang 2000; 
Takata, Chang \& Shibata 2008; Takata et al. 2010b), 
implying that  the outgoing  particles are accelerated with 
 whole gap  potential drop, whereas the incoming particles  can only receive  
$\le  10\%$ of the gap potential drop. 
Hence, we expect that the  power carried by the outgoing particles 
 is, at least,  one order of magnitude greater than 
that carried by the incoming particles. In the gap, furthermore, because 
 the curvature radiation process is 
occurring under the force balance 
 between the acceleration force and the radiation 
back reaction force, as assumed in equation~(\ref{gamma}), 
 the particles lose   most of the energy gain via the 
curvature radiation process.  Hence, the  luminosity of the outgoing 
$\gamma$-rays  is, at least,  one order of magnitude greater 
than that of the incoming $\gamma$-rays. As we will discuss 
in section~\ref{heating}, the incoming particles eventually reach the 
stellar surface and heat up the polar cap region, producing 
the observed thermal X-ray emissions. The Lorentz factor of 
the incoming particles  is reduced from $\sim 10^7$ at the inner boundary 
of the outer gap to  
$\sim 8\times 10^{5}$ at the stellar surface, indicating the X-ray emissions 
from the heated polar cap region is at least two order of magnitude 
less than the outgoing $\gamma$-rays radiated by the outgoing particles.

  As equation~(\ref{lgamma}) indicates, 
 the $\gamma$-ray luminosity from the pulsar  
depends on its spin down luminosity. On the other hand,  the spin 
down luminosity depends on the structure of  magnetosphere 
 and the inclination angle between  
 the magnetic axis and the rotation axis, indicating the predicted 
\ $\gamma$-ray luminosity from the outer gap  depends on the structure of the 
pulsar magnetosphere.   For example, Spitkovsky (2006) 
represented the dipole field structure in the ideal MHD limit, in which 
there is no accelerating field,  
 and found that the inferred spin down luminosity,  $L^{MHD}_{sd}\sim (2\pi)^4
\mu^2/(P^{4}c^3)\times (1+\sin^{2}\theta_o^2)$,  can be up to 3 times larger than 
the standard vacuum formula $L^{va}_{sd}\sim 2(2\pi)^4
\mu^2/(3P^{4}c^3)$, where  $\theta_o$ is the inclination angle.  Although 
the magnetospheric structure has not been understood well,  
the real magnetosphere with the accelerating region will be between 
the vacuum limit and the ideal MHD limit (e.g. Constantinos et al. 2011; 
Li, Spitkovsky \& Tchekhovskoy, 2011; Wada and Shibata 2011), and therefore 
the spin down  luminosity will be only
 a factor of 1-3 different from that of standard formula ($L_{sd}^{va}$). 
Furthermore, the detailed structure of the magnetosphere, 
such as the magnetic field configurations, will be more important to 
 the pulse profiles and the phase-resolved spectra, which
 are beyond the scope of this paper. 
In this paper, therefore, we apply the standard formula
 as a typical magnitude of the spin down power. 

\subsection{Heated polar cap by incoming current}
\label{heating}

Half of the particles accelerated  in the outer gap will return to the polar cap region and heat 
the stellar surface. Several rotation powered MSPs exhibit thermal X-ray radiation characterized by 
a temperature of $T\sim 10^6$~K (Zavlin 2007), which is much higher than $T\sim 10^5$~K expected 
from the standard neutrino cooling scenario (Yakovlev \& Pethick 2004 for the review). Furthermore, 
the size of the inferred emitting region corresponds to an effective radius of $10^{4-5}$~cm, which
is much smaller than the NS radius. For some MSPs the observed X-ray spectra can be fit by black body 
radiation with two components, that is, a ``core'' component described by a higher temperature 
($T_c>10^6$~K), but smaller effective radius ($R_c\sim 10^{3-4}$~cm), and a 
``rim'' component with a lower temperature ($T_r\sim 5\times 10^5$~K) and larger effective radius ($R_r\sim 10^{5}$~cm). The observed temperatures and the effective radii of the 11 $\gamma$-ray emitting 
MSPs are summarized in Table~1.  These observations provide indirect evidence for additional 
heating of the polar cap region.

It has been hypothesized that near the stellar surface, the magnetic field configuration is not dominated by a dipole field (Ruderman 1991; Chen, Ruderman \& Zhu 1998).  Higher order multipole  field configurations are likely and the strength of these components can be 1-3 orders of magnitude 
greater than the global dipole field.  The distance $\delta R_{eq}$ from the star for which the 
local magnetic field is comparable to the dipole field is estimated from the relation 
\begin{equation}
B_s\left(\frac{l+\delta R_{eq}}{l}\right)^{-3}=B_d\left(\frac{R_s+\delta R_{eq}}
{R_s}\right)^{-3}
\end{equation}
where $B_s$ is the strength of the local magnetic field at the stellar surface, $R_s$ is the stellar 
radius,  and $l\sim (1-3)\times  10^{5}$~cm is the thickness of the NS crust.  With $B_s=10-1000B_d$, 
we find $\delta R_{eq}\sim (1-3)R_s$.  We expect that because of the bending of the local magnetic 
field with a smaller curvature radius ($\sim 10^{5-6}$~cm) than $\sim 10^{7}$~cm of the dipole field,  
the curvature photons emitted between $R_s<r<R_s+\delta R_{eq}$ will illuminate a wider area on the 
stellar surface than that connected to the outer gap via the global dipolar magnetic field lines. 

\begin{figure}
\epsscale{.80}
\plotone{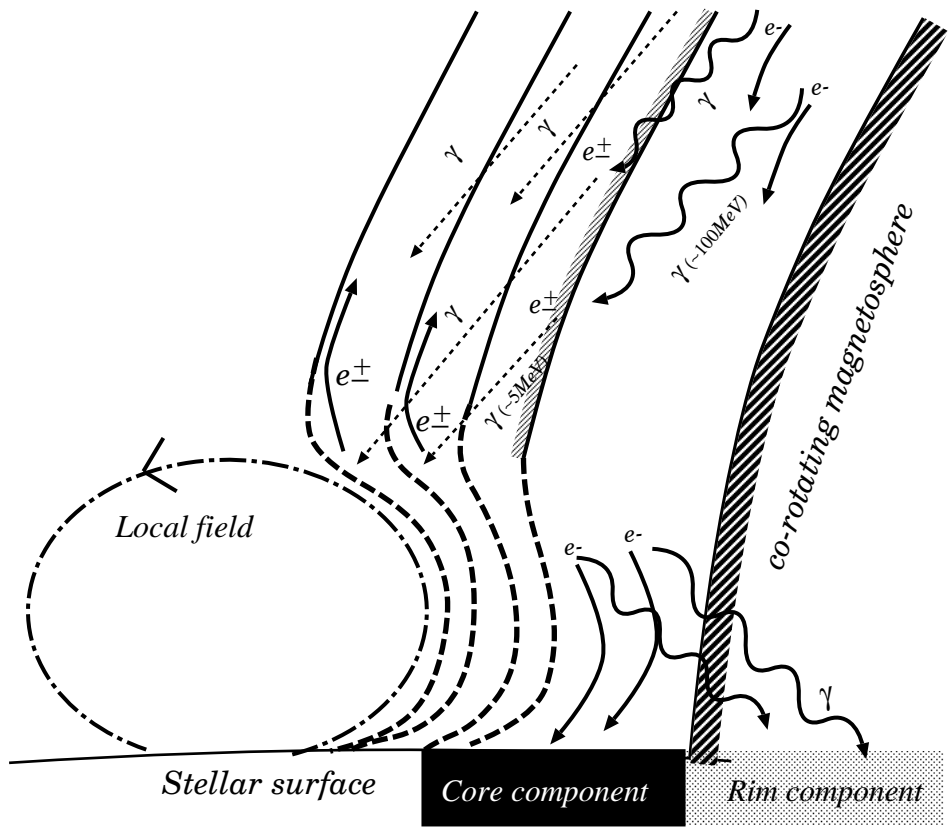}
\caption{Schematic view for the structure of the polar cap region. Near the stellar surface, the multipole 
field components dominate the global dipole field, and the trajectory of the incoming particles and the 
emission direction of the $\gamma$-rays are determined by the local field. The rim component and core 
component are produced by the irradiation of the $\gamma$-rays between $R_s\ge r\ge R_s+\delta R_{eq}$ and by 
bombardment of the incoming particles respectively.}
\label{localst1}
\end{figure}

 In Figure~\ref{localst1}, we illustrate this picture for the structure 
of the polar cap region.  The incoming relativistic particles, which 
were accelerated in the gap,   lose their energy via curvature radiation
 between the stellar surface and the inner boundary of the gap, and 
its Lorentz factor  decreases 
to $\Gamma\sim 2\times 10^{6}P^{1/3}_{-3}$ [equation~(\ref{gamma1})] 
near the stellar surface ($r\sim 2R_s$). Near the stellar surface, 
 the incoming particles  emit $\gamma$-ray photons
 with an energy $m_ec^2/\alpha_f~\sim 70$~MeV, where $m_ec^2$ is the electron 
rest mass energy and $\alpha_f$ 
is the fine structure constant (Wang et al. 1998; Takata et al. 2010b).  
These  $~70-100$MeV photons may be converted 
into pairs via the local magnetic field, whose strength 
may reach $B\sim 10^{11}$~Gauss. The local 
magnetic field will bend the trajectory of the  incoming particles so that 
the $\sim 100$~MeV photons will illuminate and heat a significant 
part of the polar cap region. The incoming particles eventually impact on 
the stellar surface with a Lorentz factor 
$\sim 8\times 10^5$ [equation~(\ref{gammas})],  and their remaining energy 
will heat an area much smaller than the polar cap region.  We expect that 
the former and latter components are observed as the rim and core components
 respectively.  With a strong local magnetic field, the typical size of 
the core component is estimated as $R_c\sim fR_p(B_d/B_s)^{1/2}$ 
where $R_p$ is the size of the polar cap region of the dipole field. 
 This yields $R_c\sim 
0.2(f/0.5)P^{-1/2}_{-3}$~km with $B_s=10^2B_d$, which is consistent with 
the observations; for example $R_c\sim 0.4$~km for PSR J~0030+4051 (Table~1).

Between the NS star surface and the inner boundary of the gap, the evolution of the Lorentz factor 
is approximately  described by 
\begin{equation}
m_ec^3\frac{d\Gamma}{dr}=\frac{2}{3}\frac{\Gamma^4e^2c}{s^2}.
\end{equation} 
Near the surface,  $s\sim\sqrt{rR_{ls}}$ provides a good approximation for the curvature radius of 
the dipole field. Assuming that $\delta R_{eq}\sim R_s$ and the Lorentz factor at $r=R_s+
\delta R_{eq}$ is much smaller than that at the inner boundary of the gap, we obtain
\begin{equation}
\Gamma(R_s+\delta R_{eq})\sim \Gamma_{1}
[\mathrm{ln}(r_0/(R_s+\delta R_{eq}))]^{-1/3}\sim \Gamma_{1},
\label{gamma1}
\end{equation}
where $r_0$ is the radial distance to the inner boundary of the gap and 
\[
\Gamma_{1}\equiv \left(\frac{R_{lc}m_ec^2} {2e^2}\right)^{1/3}\sim 2\times 10^6P_{-3}^{1/3}.
\]

In the region between $R_s<r<R_s+\delta R_{eq}$, the trajectories of the particles are described by 
the local magnetic field lines, which have a curvature radius of $\zeta \sim 10^{6}$~cm. 
In such a case, the Lorentz factor of the particles at the stellar surface can be estimated from 
\begin{equation}
\Gamma_s\sim \left(\frac{m_ec^2\zeta}{2\pi e^2}\right)^{1/3}
\sim 8\times 10^5\zeta_6^{1/3}
\label{gammas}
\end{equation}
where  $\zeta_6=\zeta/10^6$~cm. The X-ray luminosity and temperatures for the rim $(L_{X,r},T_r)$ 
and core ($L_{X,c}, T_c)$ components are calculated from  
\begin{equation}
L_{r}=(\Gamma_{1}-\Gamma_s)\dot{N}m_ec^2,~T_r=\left(\frac{L_{X,r}}{4\pi R^2_{r}
\sigma_{s}}\right)^{1/4}, 
\end{equation}
and
\begin{equation}
L_{c}=\Gamma_s\dot{N}m_ec^2,~T_c=\left(\frac{L_{X,c}}{4\pi R^2_{c}
\sigma_s}\right)^{1/4}, 
\end{equation}
respectively, where $\dot{N}$ is the number of incoming particles per unit time and  $\sigma_s$ 
is the Stefan-Boltzmann constant.  In addition, $R_{r}$ and $R_{c}$ are the effective radii of the 
heated region of the rim and core components respectively. 

Wang, Takata and Cheng (2010) fit the phase-averaged spectrum of $\gamma$-ray pulsars using the 
outer gap model and suggested that the averaged current density 
 is about 50~\% of the Goldreich-Julian value. 
Hence, we take the rate of the incoming particles  as   
\begin{equation}
\dot{N}=f\frac{c\mu}{2eR_{lc}^2}.
\label{primary}
\end{equation}
As a result, the X-ray luminosity of the rim and core components are described by 
\begin{equation}
L_{r}=(\Gamma_1-\Gamma_s)m_ec^2\dot{N}\sim 10^{32}fP_{-3}^{-5/3}\mu_{26}Q~\mathrm{erg~s^{-1}},
\label{xlr}
\end{equation}
and 
\begin{equation}
L_{c}=\Gamma_sm_ec^2\dot{N}\sim 5\times 10^{31}fP^{-2}_{-3}\mu_{26}
\zeta^{1/3}_6~\mathrm{erg~s^{-1}},
\end{equation}
where $Q=1-[(\zeta/\pi R_{lc})^{1/3}]$.  The temperature of the heated surface becomes
\begin{equation}
T_{r}\sim 2\times 10^6f^{1/4}P_{-3}^{-5/12}\mu_{26}^{1/4}R^{-1/2}_{r,5}
Q^{1/4}~\mathrm{K},
\label{trim}
\end{equation}
and 
\begin{equation}
T_{c}\sim5\times 10^6f^{1/4}P_{-3}^{-1/2}\mu_{26}^{1/4}R^{-1/2}_{c,4}
\zeta_6^{1/12}~K,
\label{tc}
\end{equation}
respectively, where $R_{r,5}=R_r/(10^{5}$~cm) and $R_{c,4}=R_c/(10^{4}$~cm).

For the canonical pulsar,   the core component 
is observed with  a luminosity  
one or two order of magnitude fainter than that of the 
rim component; for example,  $L_c\sim 10^{30}~\mathrm{erg~s^{-1}}$ and  
$L_r\sim  10^{32}~\mathrm{erg~s^{-1}}$ for the Geminga pulsar 
(Kargaltsev et al. 2005).   Halpern \& Ruderman (1993) studied 
the thermal X-ray emissions from the heated polar cap due to the bombardment of the incoming particles, which were accelerated 
in the outer gap,  and found that
the predicted luminosity $\sim 10^{32}~\mathrm{erg~s^{-1}}$ with a temperature $T> 
10^{6}$~K for the Geminga pulsar is too bright  compared with the observed luminosity of 
the core component  $L_c\sim 10^{30}~\mathrm{erg s^{-1}}$. 
Hence, they proposed that most of the X-ray photons  from the 
heated polar cap region is scattered by the resonant Compton scattering 
and are eventually redistributed as a
 thermal emission from almost entire surface with a temperature $T<10^{6}$~K, which is observed 
 as the rim component(Wang et al. 1998; Cheng \& Zhang 1999). 

For the MSPs,  the observed luminosity of the core 
component ($L_{c}\sim 10^{31}~\mathrm{erg~s^{-1}}$) 
is comparable to that of the rim component ($L_{r}$), as Table~1 indicates. 
In fact, our model predicts  that 
most of  thermal emissions  from  the 
heated polar cap regions directly emerge without the resonant scattering (section~\ref{rotation}).  
Because of the weaker  magnetic field of the MSPs, the resonant Compton scattering may be less effective as  
 compared with the case of the  canonical pulsar. 

\section{Application to the Rotation Powered Millisecond Pulsars}
\label{rotation}
As argued by Takata et al. (2010b) the outer gap can be controlled by either the photon-photon 
pair-creation process or the magnetic pair-creation process. In this section, we first review 
the outer gap model controlled by the photon-photon pair-creation process in section~\ref{photon} 
and the magnetic pair-creation process in section~\ref{magnetic}. For reference, Table~1 summarizes 
the observed parameters of the MSPs detected by the $Fermi$-LAT.

\subsection{Outer gap  controlled by the photon-photon pair-creation process}
\label{photon}
The $\gamma$-rays emitted in the outer gap can collide with the X-rays from the heated polar cap and 
convert into electron-positron pairs.  It is possible that this process, itself, controls the 
thickness of the outer gap (Zhang \& Cheng 1997, 2003). However, given the existence of the rim and core components, it is not 
clear theoretically which component controls the outer gap.  If both components illuminate the gap, 
the core component, which is of a higher temperature than the rim component, is more likely to 
control the size of the outer gap.  On the other hand, with a small effective radius of $R_c\sim 
10^{3-4}$~cm, it is possible that the core component does not illuminate the outer gap.  In this 
respect, the X-ray emission of the rim component, whose  effective radius $R_{r}\sim 10^{5-6}$~cm, 
has a greater likelihood of illuminating the outer gap.  To avoid complexity in the theoretical 
argument, therefore, we only present the case that thermal X-rays from the rim component control the 
size of the outer gap.  We remark that if the $\gamma$-ray emission is  
controlled by the core component, the predicted $\gamma$-ray luminosity is several times smaller 
than presented here.  However, the main conclusion in this paper will remain unchanged. 

The use of the pair-creation condition $E_{\gamma}E_{X}\sim (m_ec^2)^2$ together with 
equations~(\ref{lcut}) and~(\ref{trim}) leads to the fractional gap thickness controlled by the 
photon-photon pair-creation process, $f\equiv f_{p}$, as 
\begin{equation}
f_p\sim0.1P_{-3}^{26/21}\mu_{26}^{-4/7}s_*^{1/7}R_{r,5}^{2/7}Q^{-1/7}, 
\label{fp}
\end{equation}
where the typical X-ray photon energy $E_{X}=3kT_{r}$ is used. The $\gamma$-ray luminosity 
(\ref{lgamma}) and the typical  radiation energy~(\ref{lcut}) can be
 rewritten as 
\begin{equation}
L^p_{\gamma}\sim3.8\times 10^{32}P^{-2/7}_{-3}\mu_{26}^{2/7}s_*^{3/7}
R_{r,5}^{6/7}Q^{-3/7}
~\mathrm{erg~s^{-1}},
\label{plgamma}
\end{equation}
and 
\begin{equation}
E^p_{c}\sim0.8P^{3/28}_{-3}\mu_{26}^{-3/28}s_*^{-1/28}
R_{r,5}^{3/7}Q^{-3/14}~\mathrm{GeV}.
\end{equation}
The total X-ray luminosity~$L_{X}=L_{r}+L_{c}$ and the temperature of the 
heated surface are now given by 
\begin{equation}
L^p_{X}\sim10^{31}P^{-3/7}_{-3}\mu^{3/7} s_*^{1/7}
R_{r,5}^{2/7}Q^{-1/7}~\mathrm{erg~s^{-1}},
\end{equation}
\begin{equation}
T^p_{r}\sim 10^{6}P^{-3/28}_{-3}\mu_{26}^{3/28}s_*^{1/28}
R^{-3/7}_{r,5}Q^{3/14}~\mathrm{K},
\label{ptr}
\end{equation}
and 
\begin{equation}
T^p_{c}\sim3\times 10^{6}P^{-4/21}_{-3}\mu_{26}^{3/28}s_*^{1/28}R_{r,5}^{1/14}
R_{c,4}^{-1/2}\zeta_6^{1/12}Q^{-1/7}~\mathrm{K},
\label{ptc}
\end{equation}
respectively. 

 In Table~2, we compare between the observed (second-fourth columns)
and   predicted (fifth-eighth columns) X-ray and $\gamma$-ray emission properties.
For the effective radii of the rim and core components, (1) we apply the 
observational value if it is  available or (2) we apply 
the typical value $R_r=3\times 10^5$~cm and $R_c=10^4$~cm, respectively,  
if observational results are unavailable.  In addition, 
we use the curvature radii  corresponding to $s_*=0.5$ and $\zeta_6=1$.
 We find in Table~2 that the predicted surface 
temperatures of the rim ($T^p_r\sim 0.5-0.7\times 10^6$~K) and
 core components ($T^p_c\sim 1-4\times 10^6$~K) are approximately consistent
 with the observations.  The predicted $\gamma$-ray luminosity 
$L^p_{\gamma}\sim 5\times 10^{32}~\mathrm{erg~s^{-1}}$ is  consistent
 with the observations for older MSPs with 
$\tau_c> 5\times 10^{9}$~yrs, where $ \tau_c\equiv 
\dot{P}/(2P)\sim 1.5\times 10^9 P_{-3}^2/\mu_{26}^2$~yrs (Table~1). 
For the younger MSPs (PSRs J0218+4232, B1937+21 and B1957+20), 
we find that the outer gap model predicts a $\gamma$-ray 
luminosity one order of magnitude less than that from the $Fermi$ observation.

\subsection{Outer gap controlled by the magnetic pair-creation process}
\label{magnetic}
Takata et al. (2010b) argued that the incoming particles emit photons with an energy 
$m_ec^2/\alpha_f\sim 70 \rm MeV$ by curvature radiation near the stellar surface.  These photons 
can become pairs via the magnetic pair creation process and the secondary pairs can continue to 
radiate several MeV photons via synchrotron radiation. In this case, the photon multiplicity can 
easily exceed $10^4$ per incoming particle.  For a simple dipole field structure, all pairs move 
inward and cannot affect the outer gap accelerator.  However  if the local field lines near the 
surface are bent sideward due to the strong multipole field (e.g. shown in Figure~\ref{localst1}), 
the pairs created in these local magnetic field lines can have an angle greater than 90$^{\circ}$, 
which results in an outgoing flow of pairs. Only a very tiny fraction (1-10) out of $10^4$ photons 
is required to create pairs in these field lines, which are sufficient to provide screening in 
the outer gap when they migrate to the outer magnetosphere. In this model, the fractional gap 
thickness in this circumstance is
\begin{equation}
f_m\sim0.025KP^{1/2}_{-3},
\label{fm}
\end{equation}
where $K\sim B_{m,12}^{-2}s_7$ characterizes the local parameters.  Here, $B_{m,12}$ and $s_7$ are 
the local magnetic field in units of $10^{12}$G and the local curvature radius in units of $10^7$cm, 
respectively. For the case of MSPs, the local parameter can be in the range $B_{m,12}\sim 0.01-0.1$ 
and $s_7\sim 0.01-0.1$, which yields $K$ of the order of ten.  Substituting equation~(\ref{fm}) into 
equations~(\ref{lgamma}) and~(\ref{lcut}), we obtain the expected $\gamma$-ray luminosity and the typical radiation  energy as 
\begin{equation}
L^m_{\gamma}\sim 6\times 10^{33}K^{3}_{1}P^{-5/2}_{-3}\mu_{26}^2~\mathrm{erg~s^{-1}},
\label{mlgamma}
\end{equation}
and
\begin{equation}
E^m_c\sim 3K^{3/2}_{1}P^{-1}_{-3}\mu_{26}^{3/4}s_*^{-1/4}~\mathrm{GeV},
\label{ecm}
\end{equation}
respectively, where $K_1=K/10$.  The total X-ray luminosity and the temperatures of the heated 
surface are described as 
\begin{equation}
L^m_X\sim 3\times 10^{31}K_1P^{-7/6}_{-3}\mu_{26}~\mathrm{erg~s^{-1}},
\end{equation}
\begin{equation}
T^m_r\sim 10^6K_1^{1/4}P_{-3}^{-7/24}\mu_{26}^{1/4}R_{r,5}^{-1/2}
Q^{1/4}~\mathrm{K},
\label{mtr}
\end{equation}
 and 
\begin{equation}
T^m_c\sim4\times 10^{6}K_1^{1/4}P^{-3/8}_{-3}\mu_{26}^{1/4} R_{c,4}^{-1/2}\zeta_6^{1/12}~\mathrm{K},
\label{mtc}
\end{equation}
respectively.

The X-ray and 
$\gamma$-ray emission properties predicted by 
the outer gap model controlled by the magnetic pair-creation 
process near the stellar surface are summarized in ninth-twelfth columns of 
Table~2.  By comparing between the  emission properties predicted by the 
 photon-photon pair-creation model and the magnetic pair-creation model 
in Table~2, we find that the  X-ray/$\gamma$-ray emission properties
 for the older MSPs  do not  depend strongly  
on the specific pair-creation process controlling 
the outer gap. For younger MSPs (PSRs J0218+4232, B1937+21 and B1957+20), 
however, the magnetic pair-creation model predicts $\gamma$-ray 
emission about one order of magnitude
 brighter than that of the photon-photon pair-creation 
model in closer agreement with the $Fermi$ observation.

For PSR~J1614-223 (or J0437-4718), the theoretical predictions
in Tables~2 are found to lie below (exceed) the measured value.  
This  discrepancy may be affected by the local structure,
since the $\gamma$-ray luminosity described by equation~(\ref{mlgamma})
is sensitive to the local structure as $L_{\gamma}\propto K^3$. Furthermore  
the uncertainties of the solid angle and the distance may also 
contribute to these discrepancies. As indicated in Table~1,
 the observed flux of  
PSR~J1614-223 implies a $\gamma$-ray emission efficiency 
of $L_{\gamma}/L_{sd}\sim 1$
assuming a solid angle $\Delta\Omega_{\gamma}=4\pi$ and distance $d=1.3$~kpc. Comparing with 
the expected efficiency  $L_{\gamma}/L_{sd}\sim 0.1$ for the outer gap model, the 
actual solid angle and/or distance may  be several factors smaller  than 
$\Delta\Omega_{\gamma}=4\pi$ and/or $d=1.3$~kpc.  In addition to the local effect and 
the observational parameters, the observed inefficient $\gamma$-ray luminosity of  PSR~J0437-4718
may be a result of an unfavorable viewing angle. 
Takata et al. (20011a,b,c) showed that the observed $\gamma$-ray flux from 
the outer gap decreases with decreasing viewing angle as measured from 
the spin axis.  Hence, $Fermi$  has preferentially discovered pulsars 
with larger  viewing angle $\xi\sim 90^{\circ}$. For the 
$\gamma$-ray pulsars with $\xi\sim 90^{\circ}$, the 
 luminosity inferred from the flux  can be characterized 
by equation~(\ref{mlgamma}). On the other hand, if the viewing angle 
is much smaller than $90^{\circ}$, the 
 $\gamma$-ray luminosity inferred from the observed 
flux will lie below the  prediction~(\ref{mlgamma}). 
In fact, a smaller Earth viewing geometry  is preferred to reproduce 
the observed single pulse  profile of PSR~J0437-4718 (Abdo et al. 2009b)
 using  the outer gap model (Takata et al. 2011c).

\begin{figure}
\epsscale{.80}
\plotone{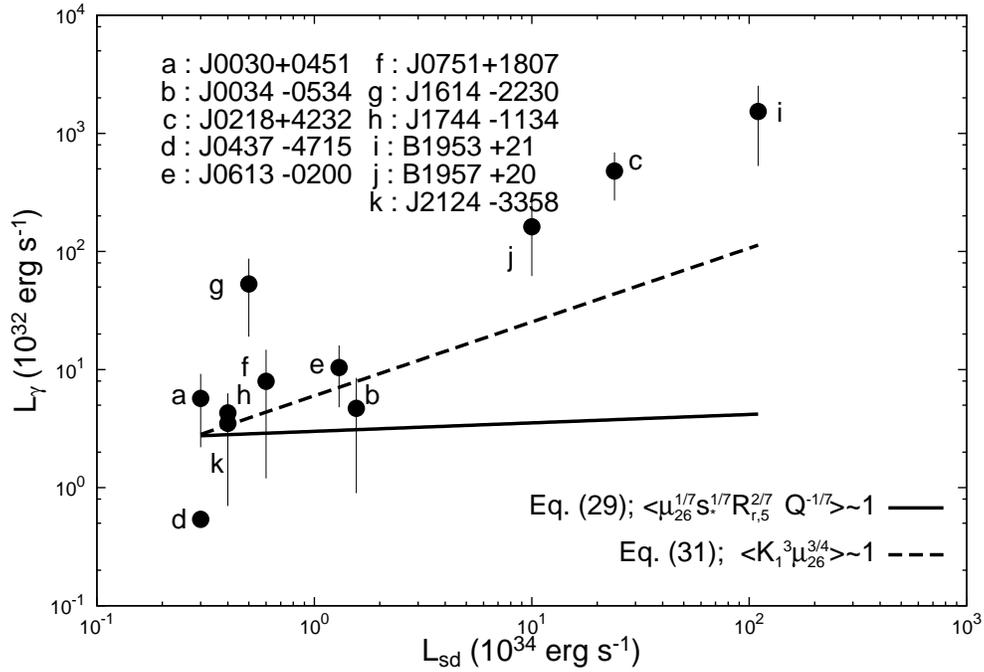}
\caption{$\gamma$-ray luminosity, $L_{\gamma}$ vs. spin down power, $L_{sd}$, for the rotation powered  MSPs. 
The outer gap model controlled by the photon-photon pair-creation process between the $\gamma$-rays and 
X-rays from the heated polar cap (solid line) and by the magnetic pair-creation process near the stellar 
surface (dashed line) respectively. The data are taken from Abdo et al. (2010).} 
\label{lgls}
\end{figure}

\begin{figure}
\epsscale{.80}
\plotone{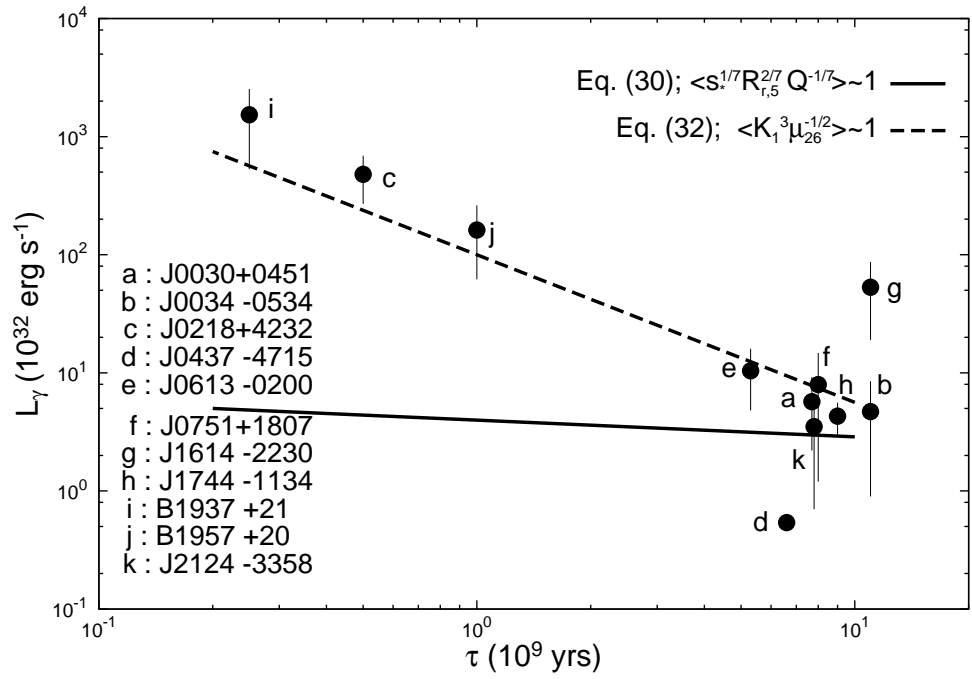}
\caption{$\gamma$-ray luminosity, $L_{\gamma}$, vs. characteristic age. The lines and dots are same as 
Figure~\ref{lgls}.}
\label{lgtau}
\end{figure}

Based on statistical grounds, Takata et al. (2010b) suggest that the outer gap controlled by the 
pair-creation model may provide a preferable explanation for the possible observational correlation 
between the characteristics of the $\gamma$-ray emission and the pulsar characteristics.  For the 
MSPs, the $\gamma$-ray luminosity $L_{\gamma}$ (\ref{plgamma}) and~(\ref{mlgamma}) can be cast in 
terms of the spin down power $L_{sd}= 2(2\pi)^4 \mu /(3c^3P^4)$ or the 
characteristic age 
$\tau_c=P/2\dot{P}$ yielding 
\begin{equation}
L^p_{\gamma}\sim 3\times 10^{32}L_{sd,34}^{1/14}\mu_{26}^{1/7}
s_*^{1/7}R_{r,5}^{2/7}Q^{-1/7}
~\mathrm{erg~s^{-1}},
\label{lglsp}
\end{equation}
or 
\begin{equation}
L^p_{\gamma}\sim 4\times 10^{32}\tau_{c,9}^{-1/7}
s_*^{1/7}R_{r,5}^{2/7}Q^{-1/7}~\mathrm{erg~s^{-1}}
~\mathrm{erg~s^{-1}},
\label{lgtaup}
\end{equation}
for the outer gap controlled by the photon-photon pair-creation process, and 
\begin{equation}
L^m_{\gamma}\sim 6\times 10^{32}L_{sd,34}^{5/8}K_1^{3}\mu_{26}^{3/4}
~\mathrm{erg~s^{-1}},
\label{lglsm}
\end{equation}
or 
\begin{equation}
L^m_{\gamma}\sim 10^{34}\tau_{c,9}^{-5/4}K_1^{3}\mu_{26}^{-1/2}
~\mathrm{erg~s^{-1}},
\label{lgtaum}
\end{equation}
by the magnetic pair-creation process.  Here, $L_{sd, 34}=(L_{sd}/10^{34}~
\mathrm{erg~s^{-1}})$ and $\tau_{c,9}=(\tau_c/10^{9}~\mathrm{yrs})$. In Figures~\ref{lgls} 
and~\ref{lgtau}, the model predictions given by equations~(\ref{lglsp})-(\ref{lgtaum}) are plotted 
with the solid lines (for the photon-photon pair-creation) or dashed lines (for the magnetic 
pair-creation).  The filled circles represent the MSPs detected by the $Fermi$-LAT.  Notwithstanding 
the large observational errors, the data points at large $L_{\gamma}$ in Figures~\ref{lgls} 
and~\ref{lgtau} may suggest that the magnetic pair-creation model is  preferred over the 
photon-photon pair-creation model for the $L_{\gamma} - L_{sd}$ and $L_{\gamma}- \tau_c$ relations.

Given that the X-rays from the heated polar cap may be prevented from illuminating the outer gap by 
the resonant cyclotron scattering  process, the magnetic pair-creation process may be the more 
important process to control the outer gap.  The cross section for Thomson scattering can be 
represented as (Halpern \& Ruderman 1993; Zhang \& Cheng 1997),
\begin{equation}
\sigma=\sigma_T(\hat{\varepsilon}\cdot\hat{B})^2+\frac{2\pi^2e^2}{m_ec}
(\hat{\varepsilon}\times\hat{B})^2\delta(\omega_B-\omega),
\end{equation}
where $\hat{\varepsilon}$ is the (electric field) polarization of the X-ray photons, $\hat{B}$ is 
the unit vector of the direction of the background magnetic field and $\omega_B=eB(r)/mc$. In 
the case where $\hat{\varepsilon}\cdot\hat{B}=0$ and for a local dipole magnetic field $B_s\sim 
10^{11}$~G, the resonant scattering will be efficient at $\delta R\sim l(B_se\hbar /E_Xm_ec)^{1/3}
\sim 7\times 10^{5}\mathrm{cm} (l/3\cdot 10^5~\mathrm{cm})(B_s/3\times 10^{11}~\mathrm{G})^{1/3} 
(T/10^{6}~ \mathrm{K})^{-1/3}$ from the stellar surface, where $l$ is the thickness of the crust. 
Since we obtain 
\begin{equation}
\int \sigma dr\sim \frac{2\pi^2 e^2}{m_ec\omega}\left(\frac{eB_sl^3}{m_ec\omega}\right)^{1/3}\sim 10^{-13}\left(\frac{l}{3\cdot 10^{5}\mathrm{cm}}\right)
\left(\frac{B_s}{3\cdot 10^{11}\mathrm{G}}\right)^{1/3}
\left(\frac{T}{10^{6}\mathrm{K}}\right)^{4/3}~\mathrm{cm^{3}},
\end{equation}
a number density $n_{\pm}>10^{13}~\mathrm{cm^{-3}}$ at $\delta R\sim 10^{6}$~cm leads to an optically 
thick cyclotron resonant scattering layer. The number density of the incoming primary particles near 
the stellar surface, which may be  about 50~\% of the Goldreich-Julian value (e.g. 
equation~\ref{primary}), becomes $n\sim B/(2Pce)\sim 3\times 10^{9} (B/10^8~\mathrm{G})P^{-1}_{-3}
~\mathrm{cm^{3}}$.  This value implies that the cyclotron resonant scattering can become optically 
thick for the multiplicity of the incoming particles due to the magnetic pair-creation ($\ge 10^{4}$, 
see Takata et al. 2010b).  Because the radial distance to the inner boundary of the outer gap from the 
stellar surface is only  $\sim 5\times 10^{6}-10^{7}~$cm for MSPs, it is possible that the scattering 
layer  prevents the illumination of the inner part of the outer gap  by X-ray photons. 
In such a case, the magnetic pair-creation process can control the outer gap.

\section{Application to Millisecond Pulsar in Quiescent LMXBs}
\label{accretion}
\subsection{Thermal X-ray emissions from MSPs}
Observations of the optical modulation of LMXBs in the quiescent state have provided indirect evidence 
for additional heating of the companion star, possibly due to the rotational energy loss of the NS 
associated with pulsar activity, e.g. J102347.68+003841.2 (Thorstensen \& Armstrong 2005), SAX 
J1808.4-3658 (Burderi et al. 2003; Deloye et al. 2008), XTE 1814-338 (D'Avanzo et al. 2009) and IGR 
J00291+5934 (Jonker, Torres \& Steeghs 2008). Specifically, the 
irradiation luminosity required to produce the amplitude of the modulation 
is $\sim 10^{33-34}~\mathrm{erg~s^{-1}}$ for the isotropic radiation.  
This is significantly greater than that associated with the X-ray 
emission from the disk or neutron star, indicating the need for the 
operation of an additional heating source.  If the pulsar magnetosphere is sufficiently clear of matter during the quiescent 
state of the LMXB, the outer gap accelerator can be activated and 
the emitted $\gamma$-rays may irradiate and heat the companion star.
If the optical modulation is a result of irradiation from the outer gap, 
the actual irradiated luminosity may be several factors less than that for the 
isotropic case because the outer gap emission is beamed with a solid angle 
$\Delta\Omega_{\gamma}\sim2-3$~radian (Takata et al. 2010b).  The amplitude of the optical 
modulation is estimated  as $L_{o}\sim (\pi \theta^2/\Delta\Omega_{\gamma})
L_{\gamma}$, where $\theta$ is the angle of the size of the 
companion star measured from the pulsar and 
$\Delta\Omega_{\gamma}$ is the solid angle of the $\gamma$-ray beam. 
If the companion fills its Roche lobe, 
$\theta\sim 0.462[q/(1+q)]^{1/3}$ for typical LMXB  $0.1<q<0.8$,  where $q$ 
is the mass ratio of the system (Frank, King \& Raine, 2002).

Deep X-ray observations have been carried out during the quiescent state to search for the thermal 
emission from the NS star surface.  The detected emissions indicate that the NS is hotter in 
comparison to expectations based on traditional cooling curves of NSs. An explanation for this 
difference is a consequence of heating associated with nuclear fusion in the crust. In this 
picture, the base of the accreted matter in the crust is sufficiently compressed by the overlying 
weight of newly accreted matter, leading to pycnonuclear reactions at $\rho\sim 10^{12-13}~\mathrm{g~
cm^{-3}}$.  These reactions release about 1-2~MeV per accreted baryon, resulting in heating of the 
crust and the core (Brown, Lars \& Rutledge 1998; Haensel \& Zdunik 1990, 2003). On a time scale of 
$10^4$~yr, thermal equilibrium is established between heating during the accretion stage and 
cooling during the quiescent stage (Colpi et al. 2001). Accordingly, the NS core and surface 
temperatures can reach $\sim 10^{8-9}$~K in the interior and $\sim 10^{6}$~K at the surface. Given this 
surface thermal emission, the outer gap in the quiescent stage may be controlled by the photon-photon 
pair-creation between the $\gamma$-rays and the X-rays from the NS surface. 

To explore the $\gamma$-ray emission from AMPs in the quiescent state, the X-ray emission in this state 
is calculated following the model description by Yakovlev et al. (2003).  The NS core 
temperature, $T_i$, surface temperature, $T_s$, and long-term time-average ($\sim 10^4$~yr) 
mass accretion rate, $<\dot{M}>$, are related by 
\begin{equation}
L_{h}(<\dot{M}>)=L_{\nu}(T_i)+L_{th}(T_s),
\end{equation}
where $L_h$ is heating term due to the  pycnonuclear reactions, $L_{\nu}$ is cooling term associated 
with neutrino emission, and $L_{th}=4\pi R_s^2\sigma_s cT_s^4$  is the thermal emission from the NS 
surface.  The heating rate is expressed as 
\begin{equation}
L_h(<\dot{M}>)=\frac{<\dot{M}>}{m_u}Q_{\nu}\sim 8.7\times 10^{33}
\frac{<\dot{M}>}{10^{-10} \msun \mathrm{yr^{-1}}}~\mathrm{erg~s^{-1}},
\end{equation}
where $m_u$ is the atomic mass unit and $Q_{\nu}\sim 1.45$~MeV is the nuclear energy release per baryon. 
The neutrino emission $L_{\nu}$ is calculated using equation~(4) in Yakovlev et al. (2003). A $T_i-T_s$ 
relation was  obtained by Gudmundsson, Pethick, \& Epstein (1983) as  
\begin{equation}
T_i\sim1.288\times 10^8(T^4_{s,6}/g_{14})^{0.455}~K, 
\end{equation}
where $g_{14}$ is the surface gravity $g=GM\mathrm{e}^{-\Phi}/R^2$ in units of $10^{14}~\mathrm
{cm~s^{-2}}$. Here, $\Phi= \sqrt{1-2GM/c^2R_s^2}$ and $T_{s,6}= T_s/10^6$~K. 

\begin{figure}
\epsscale{.80}
\plotone{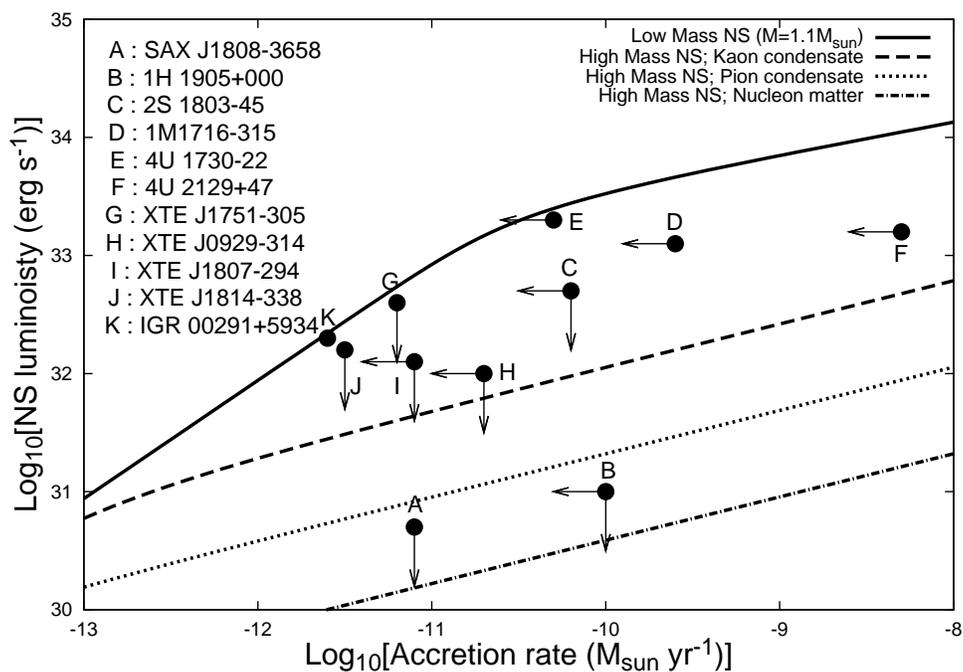}
\caption{The quiescent thermal X-ray luminosity from the NS surface as a function of the averaged accretion  rate.  The predictions are for the low mass NS (solid line) and for various enhanced cooling mechanisms of 
the high mass NS (Yakovlev et al. 2003). The observational data are taken from Heinke et al. (2009).}
\label{axplum}
\end{figure}

\begin{figure}
\epsscale{.80}
\plotone{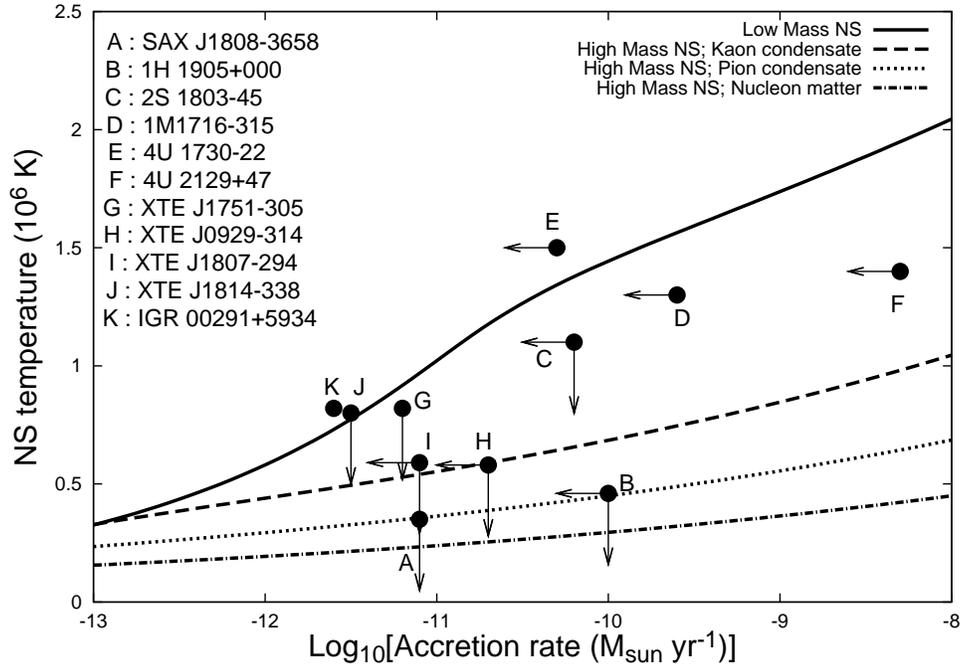}
\caption{The surface temperature of the MSP in quiescent LMXBs as a function of the averaged accretion rate. 
The lines and dots are same as Figure~\ref{axplum}.} 
\label{axptemp}
\end{figure}

Figures~\ref{axplum} and \ref{axptemp} display the X-ray luminosity and the surface temperature in 
the quiescent state as a function of the averaged accretion rate respectively. The different lines 
correspond to cooling of a low mass NS (solid line) and various enhanced cooling mechanisms for 
high mass NSs.  In Figures~\ref{axplum} and \ref{axptemp}, the observational data for various quiescent 
LMXBs are given for reference, based on the work by Heinke et al. (2009).

\subsection{$\gamma$-ray emissions from the outer gap}
\subsubsection{$\gamma$-ray luminosity}
Using the pair-creation condition $E_{\gamma}E_X=(m_ec^2)$ with $E_X=3kT$, the relation between the 
fractional gap thickness and the surface temperature is given as 
\begin{equation}
f\sim 0.12P_{-3}^{7/6}\mu_{26}^{-1/2}s_*^{1/6}T_{s,6}^{-2/3},
\end{equation}
yielding a $\gamma$-ray luminosity in the quiescent state corresponding to 
\begin{equation}
L_{\gamma}\sim 6\times 10^{32}P_{-3}^{-1/2}\mu_{-26}^{1/2}s_*^{1/2}T_{s.6}^{-2}
~\mathrm{erg ~s^{-1}}.
\end{equation}

Figure~\ref{glum} represents the predicted $\gamma$-ray luminosity for four LMXBs (SAX~J1808.4-3658, 
XTE J0929-314, XTE J1814-328 and IGR J00291-5934) as a function of the averaged accretion rate.   The 
solid line (for the low mass NS) and dashed line (for the high-mass NS with
 neucleon matter)  represent the results for the outer gap model controlled by 
the photon-photon pair-creation process between the $\gamma$-rays 
and the X-ray from the full surface cooling emissions. 
Furthermore, the dotted and dotted-dashed horizontal lines represent 
results for the outer gap model controlled  by the photon-photon pair-creation process between 
the $\gamma$-ray and the X-rays from the heated polar cap region and by the magnetic pair-creation 
process near the stellar surface respectively. Since the magnetic fields for the MSPs in quiescent 
LMXBs have not been constrained, we present the results for two extreme cases as thick and thin  
lines.  For the thick lines, we assume that $\mu_{26}=1.5P_{-3}^{7/6}$, which gives an upper limit 
of the magnetic field for the recently turned on the radio millisecond pulsar PSR~J1023+0038 (see 
Figure~\ref{pb}).  For the thin-lines, on the other hand, we estimate the dipole magnetic field from 
$\mu_{26}=P_{-3}^{7/6}/3$,  which describes the relation for the MSPs detected by $Fermi$-LAT.

It can be seen in Figure~\ref{glum} that the predicted $\gamma$-ray luminosity given by the solid and 
dashed lines increases with decreasing time averaged accretion rates. This dependence reflects the 
fact that the surface temperature decreases with a decrease of the averaged accretion rate, as 
Figure~\ref{axptemp} reveals.  In the present case, the fractional gap thickness is related with the 
surface temperature as  $f\propto T_{s,6}^{-2/3}$, indicating that the gap is thicker for lower 
accretion rates.  Since the $\gamma$-ray luminosity is expressed by $L_{\gamma}\sim f^3L_{sd}$, the 
predicted $\gamma$-ray luminosity increases with a decrease of the averaged accretion rate.  We note 
that the predicted $\gamma$-ray luminosity presented in Figure~\ref{glum} is insensitive to the 
spin periods of the known MSPs in the quiescent LMXBs because it is assumed that $\mu\propto 
P^{7/6}$, which results in $L_{\gamma}\propto P^{1/12}$.

\subsubsection{Irradiation of $\gamma$-rays to companion star}
\label{irrad}
 To explain the observed optical modulation of the companion star in quiescent state, 
Takata et al. (2010a) discussed  the irradiation of 
$\gamma$-rays from the outer gap to the companion star. The magnetospheric $\gamma$-ray irradiating the companion star may be absorbed via the so-called pair-creation process in the Coulomb field by the nuclei in the stellar matter. Further absorption will occur as the relativistic pairs created with a Lorentz factor  $\Gamma\sim 10^3$ will transfer their energy and momentum to the stellar matter via the ionization and/or the Coulomb scattering processes.  The cross section of the above pair-creation process for the photon with 
 energy $E_{\gamma}$ is given by Lang (1999) as
\begin{equation}
\sigma(E_{\gamma})=3.5\times 10^{-3}Z^2\sigma_T
\left[\frac{7}{9}\log \left(\frac{183}{Z^{1/2}}\right)-\frac{1}{53}\right]
~~\mathrm{for}~~\frac{E_{\gamma}}{2}>>\frac{m_ec^2}{\alpha Z^{1/3}}, 
\end{equation}
where $Z$ is the atomic number, $\sigma_T$ is the Thomson cross section. 
 All $\gamma$-rays irradiating the star can be absorbed 
if the column density of 
the star exceeds 
\begin{equation}
\Sigma>\frac{m_p}{\sigma}\sim 60~\mathrm{g~cm^{-2}},
\end{equation}
where we use $Z^2=3$ appropriate for the solar abundance. We can see that 
the above condition is easily satisfied for the typical low mass companion star, 
which has $\sim 0.1 \msun$ and the radius $\sim 10^{10}$~cm. 

The created pars will transfer their energy to the stellar matter via
exciting and ionizing atoms in the matter. For 
atomic hydrogen, the energy loss rate of the pairs per unit length is given 
by  Lang (1999) as
\begin{equation}
\frac{dE}{dx}\sim -2.54\times 10^{-19}N_e W~\mathrm{eV~cm^{-1}}
\end{equation}
where $N_e$ is the number density of the electrons in the matter, and $W$ 
is a factor of 10-100. All energy of the created pairs will transfer to 
the stellar material if the column density of the star exceeds 
\begin{equation}
\Sigma> 70\left(\frac{E}{1\mathrm{GeV}}\right)\left(\frac{W}{100}\right)^{-1}
~\mathrm{g~cm^{-2}},
\end{equation}
which is easily satisfied for the typical companion star.  Hence, all of 
the irradiation energy from the outer gap will transfer to the stellar matter.

In Figure~\ref{glum}, the observational data represents the
 lower limit of the irradiating luminosity $L_{irr}$ required to explain the optical modulation. We find that if the optical modulation 
originates from the irradiation of 
the $\gamma$-rays from the outer gap of the pulsar magnetosphere, the observations can constrain 
the theoretical model. For example, the level of the inferred irradiation luminosity $L_{irr}\sim 
10^{34}~\mathrm{erg s^{-1}}$ of SAX~J1808.4-3658, XTE~J1814-325 and IGR~J00291+5934 suggests  that the outer gap with the high-mass NS model is preferable.  The outer gap model with low mass NS 
predicts, on the other hand, a lower $\gamma$-ray luminosity by an order of magnitude due to the 
higher NS surface temperatures and may be applicable to J0929-314. 

It should be noted that $L_{irr}\sim 10^{34}~\mathrm{erg~s^{-1}}$ cannot be produced by the outer 
gap model controlled by the photon-photon pair-creation process between 
the $\gamma$-rays and the X-rays from the heated polar cap region. 
Within the context of the present treatment for the magnetic field determination, that is $\mu_{26}=\kappa P_{-3}^{7/6}$, where $\kappa$ 
is the proportional factor, the $\gamma$-ray luminosity described by equation~(\ref{plgamma}) is less 
dependent on the proportional factor $\kappa$. That is, $L_{\gamma}\propto \kappa^{2/7}$ and unrealistic 
values, say $\kappa\sim 1000$
 are required to produce $L_{\gamma}\sim 10^{34} ~\mathrm{erg~s^{-1}}$. However, such a model would not be consistent with the 
properties of the quiescent X-ray emission. 
On the other hand, $L_{\gamma}\propto \kappa^2$ for the outer 
gap with the magnetic pair-creation process implies that such a model can produce $L_{\gamma}\sim 10^{34}~\mathrm{erg~s^{-1}}$ with a 
reasonable value of the NS magnetic field. 
As indicated by the dashed-dotted lines in Figure~\ref{glum} 
(also Figure~\ref{xlum}), the model prediction is sensitive to the specific relation between the spin 
period and  magnetic moment. Hence a measurement  the possible $P-\mu$ relation for the accreting 
MSPs is desired for further discussion on the origin of $\gamma$-ray emission from the MSPs in quiescent LMXBs.

\section{Magnetospheric Non-thermal X-ray emissions from MSPs }
\label{nonx}
\subsection{Rotation powered MSPs}
The $\gamma$-rays ($>$GeV) from the outer gap may be converted into pairs via the photon-photon 
pair-creation process involving the thermal X-rays from the NS surface before escaping the 
magnetosphere.  The secondary pairs produced near the light cylinder, where the magnetic field 
$B\sim 10^{5-6}$~G, will emit non-thermal X-rays via the synchrotron radiation process with a 
typical energy of $E_{n,X}\sim 2(\Gamma/10^3)^2(B/10^{6}~\mathrm{G})^2
(\sin\theta/0.1)$~keV. Here, 
$\Gamma_s\sim (1~\mathrm{GeV}/2m_ec^2)$ is the typical Lorentz factor of the secondary pairs, and 
$\theta$ is the pitch angle.  The synchrotron damping length of the secondary pairs
is  $l_d\sim  1.5\times 10^{6}~(\Gamma_s/10^{3})^{-1} (B\sin\theta/10^{5})^{-2}$~cm,  implying the 
secondary pairs quickly lose their perpendicular momentum inside the light cylinder. 
In such a case, the luminosity of the  non-thermal X-rays is estimated as 
\begin{equation}
L_{n,X}\sim \tau_{X\gamma}L_{\gamma},
\label{nxl}
\end{equation}
where $\tau_{X\gamma} (<1)$ is the optical depth of the photon-photon pair-creation process. Note that 
because the local cyclotron energy near the light cylinder is less than the X-ray energy, the resonant 
scattering is inefficient near the light cylinder, implying the non-thermal X-rays can freely escape 
from the magnetosphere.  

\begin{figure}
\epsscale{.80}
\plotone{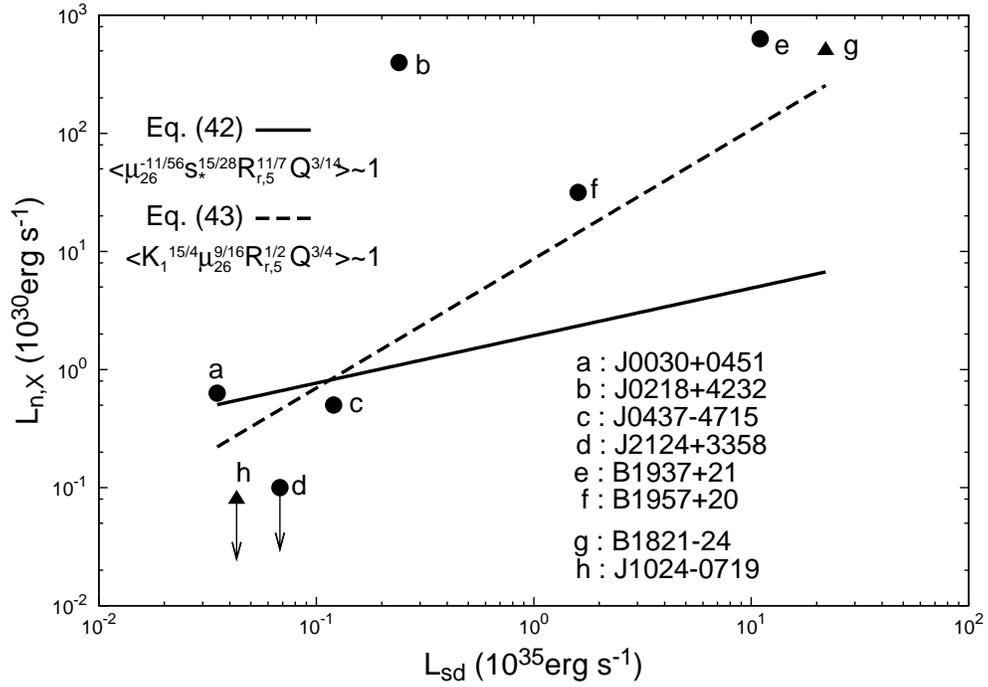}
\caption{The non-thermal X-ray emission from the magnetosphere for the rotation powered MSPs. The solid and 
dashed lines represent the results for the outer gap model controlled by the photon-photon pair-creation 
process and by the magnetic pair-creation process.  The filled circle and the triangles represent the 
$\gamma$-ray MSPs and radio MSPs respectively. The data are taken from Zavlin (2007).}
\label{millix}
\end{figure}

It can be seen from equations~(\ref{xlr})-(\ref{tc}) that the ratio of the photon 
number density, which is proportional to $L_{r}/T_r$, of the rim component is about a factor of ten 
larger than that of the core component, indicating the optical depth $\tau_{X\gamma}$ near the light 
cylinder can be estimated as 
\begin{equation}
\tau_{X\gamma}\sim \frac{L_{r}}{4\pi R_{lc}^2cE_{r}}\sigma_{X\gamma}
R_{lc}\sim0.014f^{3/4}P_{-3}^{-9/4}\mu^{3/4}_{26}R_{r,5}^{1/2}Q^{3/4}. 
\end{equation}
Here, $E_{r}=3kT_r$ and $\sigma_{X\gamma}\sim \sigma_T/3$ with $\sigma_T$ corresponding to the 
Thomson cross section. Inserting  the gap fractions described by equations~(\ref{fp}) and~(\ref{fm}) 
into equation~(\ref{nxl}), the non-thermal X-ray luminosity can be expressed as a function of the 
spin-down power as 
\begin{equation}
L_{n,X}\sim 5\times 10^{29}L_{sd,35}^{45/112}\mu_{26}^{-11/56}s_*^{15/28}R^{11/7}_{r,5}
Q^{3/14}~\mathrm{erg~s^{-1}}
\label{nxp}
\end{equation}
for the outer gap model controlled by the photon-photon pair-creation process and 
\begin{equation}
L_{n,X}\sim 6\times 10^{30}L_{sd,35}^{35/32}K_1^{15/4}\mu_{26}^{3/8}R^{1/2}_{r,5}
Q^{3/4}~\mathrm{erg~s^{-1}},
\label{nxm}
\end{equation}
by the magnetic pair-creation process.  The  model results described by equations~(\ref{nxp}) 
and~(\ref{nxm}) are plotted in Figure~\ref{millix} with the solid and dashed lines respectively. For the 
observational data in Zavlin (2007), we represent the $Fermi$ MSPs with the filled circles and the radio MSPs with 
filled triangles. We find in Figure~\ref{millix} that the outer gap model controlled by the magnetic 
pair-creation process can explain the observed possible correlation between the non-thermal X-ray 
luminosity and the spin down power slightly better than the model controlled by the photon-photon pair 
creation process. 

For the synchrotron emission from the secondary pairs, it is expected that the spectrum is described 
by a photon index of  $\alpha\sim 1.5$ below $E<E_{n,X}$  (Takata, Chang \& Cheng 2007). On the other 
hand, the secondary particles will escape from the light cylinder with a Lorentz factor  $\Gamma_{c} 
\sim 5\times 10^{2}$, for which the synchrotron damping length is comparable to the light cylinder radius.  
We  expect that the emission of the secondary  pairs beyond the light cylinder is not observed 
as pulsed emission because the co-rotation motion  with 
central star can not be retained beyond the light cylinder.
 In such a case, the observed pulsed emissions has a break 
at the energy of $E_{c}\sim 400~(\Gamma_c/5\cdot 
10^{2})^2(B/10^{6})(\sin\alpha/0.1)$~eV, below which the spectrum is characterized by  
a photon index  $\alpha=2/3$, which corresponds to the 
spectrum  described by the emission from the single particle. Hence, 
the present model predicts that the  non-pulsed X-ray emissions 
from rotation powered MSPs  are described by spectra with a photon index
 $\alpha=2/3-1.5$, which may not be in conflict  with the observed range 
$\alpha\sim 1-2$ (Zavlin 2007). Furthermore, the present model predicts that 
a spectral cut-off of the secondary emissions appears at  
$E_{n,X}\sim 2(\Gamma/10^3)^2(B/10^{6}~\mathrm{G})^2
(\sin\theta/0.1)$~keV.  If the cut-off energy position is 
located in the observation energy range, the photon index fit to the observed spectrum 
 with the single power law function can be larger than $\alpha=1.5$.

\subsection{Accretion powered pulsars in quiescent  LMXBs}
\label{lmxbnx}
\begin{figure}
\epsscale{.80}
\plotone{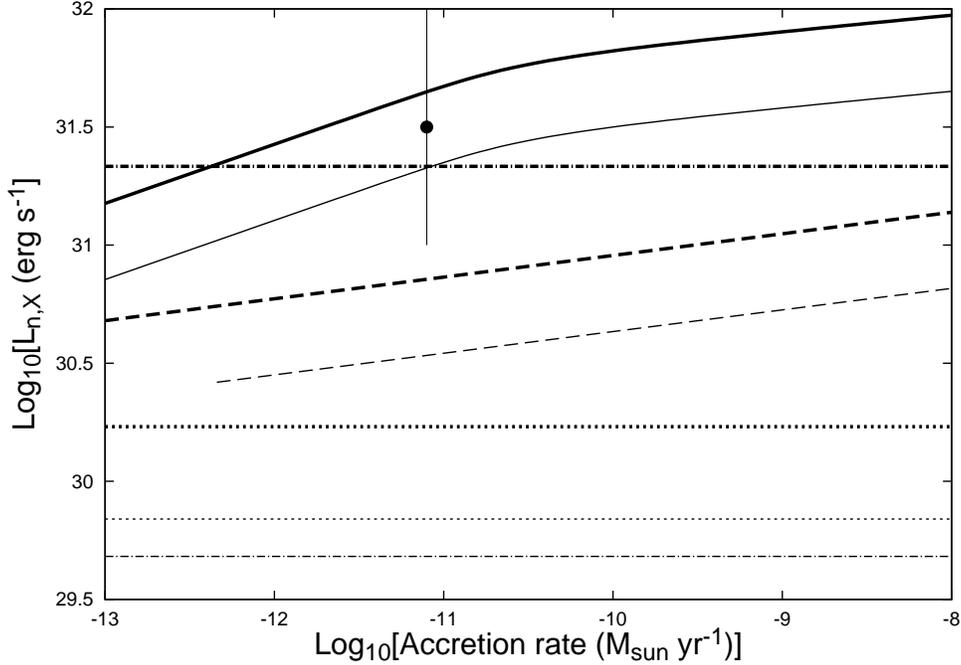}
\caption{The pulsed non-thermal X-ray emission from the MSP in a quiescent LMXB. The solid and dashed lines 
represent the results for the outer gap models controlled  by the photon-photon 
pair-creation process between the $\gamma$-rays and whole surface cooling 
X-ray emissions of a  low mass NS and  of a high-mass NS 
with nucleon matter and respectively.  For 
comparison, the dotted line and dashed-dotted line represent results for the outer gap model controlled 
by the photon-photon pair-creation process between the $\gamma$-rays and X-rays from the heated polar cap 
and by the magnetic pair-creation process near the stellar surface, respectively. The thick and thin lines 
represent the magnetic field determined from $\mu_{26}=P^{7/6}_{-3}/3$ and $\mu_{26}=1.5P_{-3}^{7/6}$ with 
$P=2.5$~ms, respectively. The filled circle represents the non-thermal emission from  SAX~J1808.4-3658 in 
quiescence.}
\label{xlum}
\end{figure}

The predicted relation between the non-thermal X-ray 
luminosity in the quiescent state and the averaged accretion rate for LMXBs is summarized in 
Figure~\ref{xlum}.  The non-thermal X-ray luminosity 
is calculated from $L_{n,X}\sim \tau_{X\gamma} L_{\gamma}$, where  the optical depth $\tau_{X\gamma}$ 
is given by $\tau_{X\gamma}=L_{th}(T_s)\sigma_{X\gamma}/(4\pi R_{lc}cE_{s})$ with $E_s=3kT_s$. 
The thermal X-ray luminosity  $L_{th}$ and the surface temperature  $T_s$ correspond to the results 
represented in  Figures~\ref{axplum} and~\ref{axptemp}, respectively.  In addition, we calculate the 
expected $\gamma$-ray luminosity $L_{\gamma}$ with $P=2.5$~ms of SAX~J1808.4-3658 for reference. Note 
that the calculated X-ray luminosity is not sensitive to the spin periods in the observed range 
of presently known MSPs in quiescent LMXBs.  In Figure~\ref{xlum}, the solid and dashed lines 
represent the non-thermal X-ray luminosity for the low-mass NS and the high mass NS models with 
nucleon matter cooling processes respectively. The thick and thin lines represent the results for  
magnetic fields determined from $\mu_{26}=1.5P_{-3}^{7/6}$ and $\mu_{26}=P_{-3}^{7/6}/3$ respectively. 
For comparison, we plot the model predictions if the outer gap is  controlled by the photon-photon 
pair-creation with the X-rays from the heated polar cap  (dotted lines) and by the magnetic 
pair-creation process near the NS surface (dashed-dotted lines) respectively. 

\begin{figure}
\epsscale{.80}
\plotone{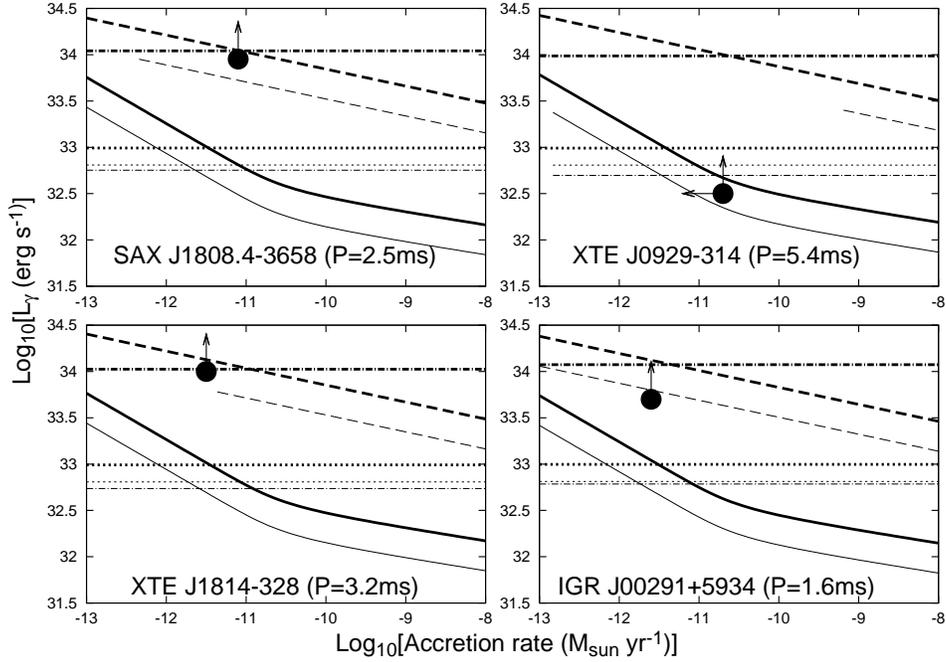}
\caption{The predicted $\gamma$-ray luminosity as a function of the averaged accretion rate.  The results for the outer gap controlled by photon-photon pair-creation 
between $\gamma$-rays and X-rays from full surface cooling emissions 
are represented by the solid line for low mass NS and by the dashed line 
for the high-mass NS with nucleon matter.  In addition,  
the dotted line and dashed-dotted line represent results for the outer gap model controlled by the 
photon-photon pair-creation process between the $\gamma$-rays and 
X-rays from the heated polar cap and by the magnetic pair-creation process
 near the stellar surface, respectively. The thick and thin lines represent 
the magnetic field determined from $\mu_{26}=P^{7/6}_{-3}/3$ and 
$\mu_{26}=1.5P_{-3}^{7/6}$, respectively. In each line, 
the fractional gap thickness is limited below  $f\le 1$.   The observational data (filled circles) represents the irradiating luminosity required to explain the optical modulation.} 
\label{glum}
\end{figure}

For the case of the outer gap controlled by the surface X-ray emission (solid and dashed lines) in 
Figure~\ref{xlum}, we find that the predicted non-thermal X-ray luminosity decreases with decreasing 
time averaged mass accretion rates, although the $\gamma$-ray luminosity increases as Figure~\ref{glum} 
shows. This results from the fact that the number of the thermal X-ray photons, which absorb the 
$\gamma$-ray photons in the outer magnetosphere, decreases with a decrease of the accretion rate, and 
the decrease of the thermal X-ray luminosity is more rapid than the increase of the $\gamma$-ray 
luminosity, as Figures~\ref{glum} and~\ref{xlum} show.  As a result, the present non-thermal X-ray 
luminosity, which is proportional to $L_{n,X}\sim \tau_{X\gamma}L_{\gamma}\propto L_{th}(<\dot{M}>) 
L_{\gamma}$, decreases with the decrease of the accretion rate.  It is also found in Figure~\ref{xlum} 
that the non-thermal X-ray luminosity is less dependent on the accretion rate in comparison with the 
$\gamma$-ray luminosity. For example, if the accretion rate changes between $10^{-13}~\msun~\mathrm{
yr^{-1}}$ and $10^{-8}~\msun \mathrm{yr^{-1}}$, the X-ray luminosity varies less than factor of ten 
(see Figure~\ref{xlum}), whereas the $\gamma$-ray luminosity varies more than factor of ten (see 
Figure~\ref{glum}). 

Finally, it has been known that X-ray emission (0.5-10~keV) 
from SAX~J1808.4-3658 in quiescence is well fit by 
a power law spectrum  with a photon index of $\alpha\sim 1.4-2$  and 
with a luminosity of $L_{n,X}\sim 10^{31-32}~\mathrm{erg~s^{-1}}$ (Heinke et al. 
2009).  Because the pulsed period in the non-thermal 
X-ray emissions has not been confirmed yet, 
its origin is still unclear.   As Figure~\ref{xlum} indicates,  
if the non-thermal X-ray emission 
originates from the magnetosphere, the outer gap model with the 
low-mass X-ray NS or with the magnetic pair-creation process 
for the gap closing may be  more favorable, although the observational 
error is large.   As we discussed  in section~\ref{irrad} 
(c.f. Figure~\ref{glum}), the optical modulation of the companion star 
observed during  the quiescent state  can be explained 
by $\gamma$-ray irradiation from the outer gap  with the high-mass 
X-ray NS  or with the magnetic pair-creation process for the gap closing. 
For the high-mass NS model, the predicted non-thermal X-ray 
luminosity associated with the outer gap  may be insufficient to explain 
the observational result, as Figure~\ref{xlum} shows, indicating the need for 
an additional component. In this case, the non-thermal emission 
originating from  the interaction between the pulsar wind and the stellar wind 
via an intra-binary shock may be observable,  
as we will discuss in  section~\ref{intra}.

\section{High-energy emissions from intra-binary shock}
\label{intra}
Stappers et al. (2003) detected the unresolved X-ray emission around a binary system  
composed of the millisecond pulsar (PSR~B1957+20) and low mass star (LMS), 
 in which the wind of the companion star eclipses
 the pulsed radio emission for $\sim 10$\% of 
every orbit. They  suggested that  
the pulsar wind is ablating the low-mass companion star, and that 
the observed unpulsed   X-ray emission 
from the PSR~B1957+20 binary system 
originates from the interaction between the pulsar 
wind and the stellar wind via an intra-binary shock 
(Arons \& Tavani 1993; Cheng, Taam \& Wang 2006). 
Similar binary systems, so called ``black widow'' systems, in which 
the MSP is destroying a low mass companion star, has also been discovered as $Fermi$ $\gamma$-ray sources 
in the Galactic field (Table~3), suggesting  the existence of high energy particles in 
the ``black widow'' system (Roberts et al. 2011).  For AMP,  the intra-binary shock in quiescent 
state is also suggested to explain the observed  non-thermal X-ray emission from  SAX~J1808.4-3658 
(e.g. Campana et al. 1998).  

The high-energy emission from an intra-binary shock between 
a pulsar and a high-mass companion star has been established 
for  so called $\gamma$-ray binary PSR~B1256/LS~2883 system, 
which is composed of a canonical 
radio pulsar with a period of $P\sim 48$~ms and 
high mass Be star~(Tam et al. 2011; Abdo et al. 2011; Aharonian et al. 2009). 
In the $\gamma$-ray binary, it has been proposed that 
the shock stands at the interface between the 
pulsar wind and the Be wind/disk, and   unpulsed 
 radio to TeV radiations are produced 
 via the synchrotron and inverse Compton processes of the accelerated
 particles at the shock (Tavani \& Arons 1997; Takata \& Taam 2009; 
Kong et al. 2011). 
The  observed photon index in the X-ray bands from the $\gamma$-ray binary 
varies with  the orbital phase in the range  $\alpha=1-2$. 

 For a pulsar/low mass star  binary, Arons \& Tavani (1993) proposed the
 intra-binary shock model  to explain the unpulsed X-ray emissions 
from the PSR~B1957+20/LMS system (see also Cheng et al. 2006).  In particular,  
the separation between the pulsar and the companion star   
is order of $a_o\sim 10^{10-11}$~cm, which is about 
2-3 orders of magnitude smaller than $a_o\sim 1$~AU 
of pulsar/high mass star system,  PSR~B1256/LS~2883. 
Therefore, if the observed non-thermal X-ray emission from  PSR~B1957+20, is produced by 
an intra-binary shock, the MSP/LMS system (i.e., black widow system) 
provides a unique laboratory to probe the physics of 
the pulsar wind in the vicinity of the neutron star.  In this section, 
therefore, we  discuss the emission from the intra-binary shock in 
pulsar/LMS systems, and  examine the predicted $\gamma$-ray and 
X-ray luminosity for known  black widow systems. 
 
It has been pointed out that the synchrotron 
emission from the shock caused by the interaction between 
the pulsar wind and ISM contributes to the resolved X-ray nebula  
around PSR~1957+20 (Stappers et al. 2003; Huang \& Becker 2007; 
Cheng et al. 2006). On the other hand, we expect that the emission 
from the pulsar wind nebula does not extend to $\gamma$-ray bands, 
and therefore is  not a candidate for the origin of 
the $\gamma$-ray emission detected by the $Fermi$ observations. 
The  Lorentz factor of the accelerated particles at 
the shock between pulsar wind 
and ISM may be limited below the critical value, at which the  gyroradius,
 $r_g=\Gamma m_ec^2/eB$, is equal to the size of the shock. Assuming 
the shock size $r_s\sim 5\times 10^{16}~$cm and the magnetic 
field $B\sim 10^{-5}$~G (Cheng el al. 2006), the Lorentz factor is 
limited below $\Gamma=3\times 10^{8} (B/10^{-5}\mathrm{G})
(r_s/5\cdot 10^{16}\mathrm{cm})$. This indicates that 
 the synchrotron spectrum extends  up to the hard X-ray bands, $E_c\sim 
(3h\Gamma^2 eB/4\pi m_ec)\sim 15(B/10^{-5}\mathrm{G})^3(r_s/5\cdot10^{16}
\mathrm{cm})^2$keV, implying the spectrum can not extend in $\gamma$-ray energy bands.  We, henceforth, examine  the high-energy emission
 from an intra-binary shock and from the  magnetosphere in black widow systems as possible sites for the origin 
of the $\gamma$-ray emissions indicated by the $Fermi$ observations.

The distance  ($r_s$) from the pulsar to the intra-binary shock
may be  estimated by the pressure 
balance as 
\begin{equation}
\frac{L_{sd}}{4\pi r_s^2 c}=\frac{\dot{M}v_g}{4\pi f_{\omega}(a_o-r_s)^2},
\label{shock}
\end{equation}
where we assumed that the pulsar wind is emitted with the  solid angle $4\pi$, 
 $\dot{M}$ is the mass loss rate from the star, $v_g$ is the velocity for 
gaseous material, and $f_{w}$ is the outflow fraction in units of $4\pi$.  
For the black widow system, PSR~B1957+20, the mass loss rate from LMS 
is expected as $\dot{M}_c\sim M_{c}\dot{P}_{b}P_b^{-1}
\sim 10^{-10}M_{\odot}~\mathrm{yr^{-1}}$, where $M_{c}=0.02M_{\odot}$ is the mass of 
LMS,  $P_{b}=33001$~s is the orbital period and $\dot{P}_b\sim 10^{-11}$ 
is the orbital period derivative  (Fruchter et al. 1990).
For the quiescent LMXBs, the mass loss rate  will be smaller than 
$\dot{M}\sim 10^{-11}-10^{-12}M_{\odot}~\mathrm{yr^{-1}}$,  based on the study 
by Heinke et al. (2009).

With the mass-loss rate of 
$\dot{M}\sim 10^{-10}-10^{-12}M_{\odot}~\mathrm{yr^{-1}}$, 
we find that  the shock distance from the pulsar
 is of the order of the orbital separation. For example, 
if we assume the stellar 
wind velocity is the escape velocity from the system, 
we obtain $v_g\sim \sqrt{2GM_{\odot}/a_o}\sim 10^{8}~\mathrm{cm~s^{-1}}$, where 
$a_0\sim5\times 10^{10}$~cm is the typical separation between two components. 
Using   $L_{sd}=10^{34}~\mathrm{erg~s^{-1}}$, 
$\dot{M}=10^{-11}M_{\odot}\mathrm{yr^{-1}}$, $a_0=5\times 10^{10}~$cm, and
 $f_{\omega}=1$, the equation~(\ref{shock}) implies 
the shock distance of $r_s\sim 3.5\times 10^{10}~$cm. Hence, 
 the shock radius from the pulsar  will be 
of order of the separation between two components.

We assume that the kinetic energy dominated flow of the pulsar wind 
is formed within the distance $r\sim a_0$ from the pulsar, that is, 
the so called magnetization parameter  $\sigma$, which is the  
ratio of the magnetic energy to kinetic energy of the unshocked flow, 
is smaller than unity, although the formation of the kinetic energy 
dominated flow within  $r\sim 10^{10-11}$~cm from the MSP is still of matter of 
debate (Lyubarsky \& Kirk 2001).  The magnetic field upstream and 
behind the shock are estimated as $B_1(r)=(L_{sd}\sigma/r^2c)^{1/2}$ and 
 $B_2(r_s)=3B_1(r_s)$, respectively.   
We assume a power law distribution of the accelerated particles at the shock,  
that is, $f(\Gamma)\propto \Gamma^{-p}$ for  $\Gamma_{min}\le\Gamma
\le\Gamma_{max}$.  The photon index in the 0.5-10~keV band 
is observed as  $\alpha\sim 2$ for 
 PSR~B1957+20 (Huang \& Becker 2007)  and 
 $\alpha=1.4-2$ for SAX~J1808.4-3658 (Heinke et al. 2009), implying  
a power law index  $p=1.8-3$, which is found in  the range predicted by 
the shock acceleration model (Baring 2004). We assume 
that the minimum Lorentz factor is comparable to the Lorentz factor 
of the bulk motion of the un-shocked flow, which is 
 $\Gamma_{min}\sim 10^{5}$ for $\sigma\ll 1$ (Takata \& Taam 2009). 
We determine the maximum Lorentz factor of the accelerated particles as 
$\Gamma_{max}=\mathrm{Min}(\Gamma_{syn},\Gamma_g)$, where 
$\Gamma_{syn}$ is the Lorentz factor at which 
the synchrotron cooling time scale 
$\tau_s\sim9m_e^3c^5/4e^4B^2\Gamma$ is equal  
 to the acceleration time scale, $\tau_{acc}\sim \Gamma m_ec/eB_2$, that is,  
 $\Gamma_{syn}\sim 5\times 10^7L_{sd,34}^{-1/4}\sigma_{0.1}^{-1/4}r^{1/2}_{s,11}$,
 where $L_{sd,34}=(L_{sd}/10^{34}~\mathrm{erg~s^{-1}})$, 
$\sigma_{0.1}=(\sigma/0.1)$ and $r_{s,11}=(r_s/10^{11}\mathrm{cm})$. In addition, 
  $\Gamma_g$ is the Lorentz factor at which the gyroradius is equal 
to the size of the system ($\sim r_s$), that is, $\Gamma_g\sim 3\times 10^{8}
L_{sd,34}^{1/2}\sigma_{0.1}^{1/2}$. 
The synchrotron spectrum extends from 
$E_{min}\sim 260L_{sd,34}^{1/2}\sigma_{0.1}^{1/2}r_{s,11}^{-1}$~eV to soft-$\gamma$-ray bands  $E_{max}\sim 2
m_ec^2\sin\theta/8\alpha_f\sim 200~$MeV if $\Gamma_{max}=\Gamma_{syn}$, 
where $\alpha_f$ is the fine structure constant, and $E_{max}\sim 
 35~L_{34,sd}^{3/2}\sigma_{0.1}^{3/2}r_{s,11}^{-1}$~GeV if $\Gamma_{max}=\Gamma_{g}$.

The  steady 
continuity equation, $\partial \dot{\Gamma}N_{tot}(\Gamma)/\partial
 \Gamma=\dot{Q}$, implies 
that the distribution of the total number within the radiation cavity is 
expressed as $N_{tot}(\Gamma)\propto  \Gamma^{-p}$ for $\Gamma_{min}
\le \Gamma\le \Gamma_c $ in the slow cooling regime 
and $N_{tot}(\Gamma)\propto \Gamma^{-1-p}$  for $\Gamma_{c}\le \Gamma\le 
\Gamma_{max}$ in the fast cooling regime, where $\Gamma_c$ is the Lorentz factor 
at which the synchrotron cooling  scale $\tau_{s}$ is equal to the dynamical 
time scale $\tau_{d}\sim r_s/c$ (see Kong et al. 2011).
 In addition,  the normalization is determined 
 from the energy conservation that $\int \Gamma N(\Gamma)d\Gamma=\eta L_{sd}
\tau_d/m_ec^2$ (Arons \& Tavani 1993), where $\eta$
 is the solid angle that the pulsar wind is stopped by the injected material 
from the low mass star. For the case of PSR~B1957+20, the eclipse of 
the pulsar ($\sim$10-20\% of the orbital period) suggests that the angle of the 
mass flow from the LMS measured from the pulsar is $\theta_a\sim 0.2$. 
If the pulsar wind is emitted spherically, 
then $\eta\sim \pi\theta_a^2/4\sim 0.03$.  

The photon index of the emission 
is characterized by $\alpha=(p+1)/2$ for $E_{min}<E<E_c$ and 
by $\alpha=(p+2)/2$ for $E_c<E<E_{max}$, where $E_c\sim 1.6L_{sd,34}^{-3/2}\sigma_{0.1}^{-3/2}r_{s,11}$~MeV is the characteristic  
photon energy emitted by the particles with the Lorentz factor $\Gamma_c$. 
The predicted luminosity in the $E_1<E<E_2$  energy band is obtained by 
\begin{eqnarray}
&&L_{s}(E_1<E<E_2))\sim\int_{\Gamma_{min}}^{\Gamma_{max}}\int_{E_1}^{E_2}N_{tot}P_{syn}dEd\Gamma \nonumber \\
&\sim&2\times 10^{30}\left(\frac{\eta}{0.03}\right)
\left(\frac{\Gamma_{c}}{10^5}\right)L^2_{sd,34}
\sigma_{0.1}r_{s,11}^{-1}Q
~\mathrm{erg~s^{-1}} \mathrm{~~for~~} p>1,
\label{inx-ray}
\end{eqnarray}
 where $P_{syn}$ is the synchrotron power per unit energy. In addition, 
the factor  $Q$ is  expressed as 
 \[
Q=(1-p)(2-p)\frac{[\int_{\Gamma_{min}/\Gamma_c}^{1}\Gamma'^{2-p}d\Gamma'+
\int_{1}^{\Gamma_{max}/\Gamma_{c}}\Gamma'^{1-p}d\Gamma']
\int_{x_{min}}^{x_{max}} F(x)dx}{(1-p)[1-(\Gamma_{min}/\Gamma_{c})^{2-p}]-(2-p)
[(\Gamma_{max}/\Gamma_{c})^{1-p}-1]}
\]
where $\Gamma'=\Gamma/\Gamma_{min}$ and $x=E/E_s$ with 
$E_s=3h\Gamma^2eB\sin\theta/4\pi m_ec$, and $x_{min}=E_1/E_s$
 and $x_{max}=E_2/E_s$ and $F(x)=x\int_x^{\infty}K_{5/3}(y)dy$ with 
$K_{5/3}$ being the modified Bessel function of order 5/3. 

For SAX~J1808.4+3365, the spin down power  is expected  to be 
$L_{sd}\ge  10^{34}~\mathrm{erg~s^{-1}}$  from the enhancement of 
the optical emissions, and the separation between two components is 
inferred  as $a_0\sim 3\times 10^{10}$~cm (Chakrabarty \& Morgan, 1998). 
Assuming $L_{sd}=5\times 10^{34}~\mathrm{erg~s^{-1}}$, for example, 
 we find that  
the expected X-ray luminosity in the 0.5-10~keV energy band from equation
~(\ref{inx-ray}) can be consistent with the observation, 
$L_{X}\sim 10^{31-32}~\mathrm{erg~s^{-1}}$ (Heinke et al. 2009), when the 
magnetization parameter is larger than $\sigma\sim10^{-2}$.

For the black widow systems, the predicted luminosity~(\ref{inx-ray}) 
in the 0.5-10~keV and 0.1-10~GeV energy bands are  summarized   
in the seventh and eighth columns in Table~3, respectively, where the first 
and the second values corresponds to the results for  the 
power law index $p=1.5$ and $p=3$, respectively. In addition,  we assume 
 the separation between two components (sixth column) 
 as the shock distance ($r_s$). In Table~3, 
we  also present  the   $\gamma$-ray (ninth  column)
 and X-ray (tenth column) luminosity predicted by  the outer gap model
 controlled by the magnetic pair-creation process.  

The predicted luminosity of the intra-binary shock emission 
depends on the power law index ($p$) and magnetization parameter ($\sigma$). 
In Table~3, we see that the  $\gamma$-ray luminosity  predicted by 
the intra-binary shock changes by at least  two orders of magnitude with  
the power law index between $p=1.5$ and $p=3$. This dependence of the $\gamma$-ray 
luminosity on the power law index is more sensitive than 
that for the X-ray luminosity, since 
the number of particles that emit GeV photons  is very sensitive to 
the photon index,  while most particles emit X-rays.

\begin{figure}
\epsscale{.80}
\plotone{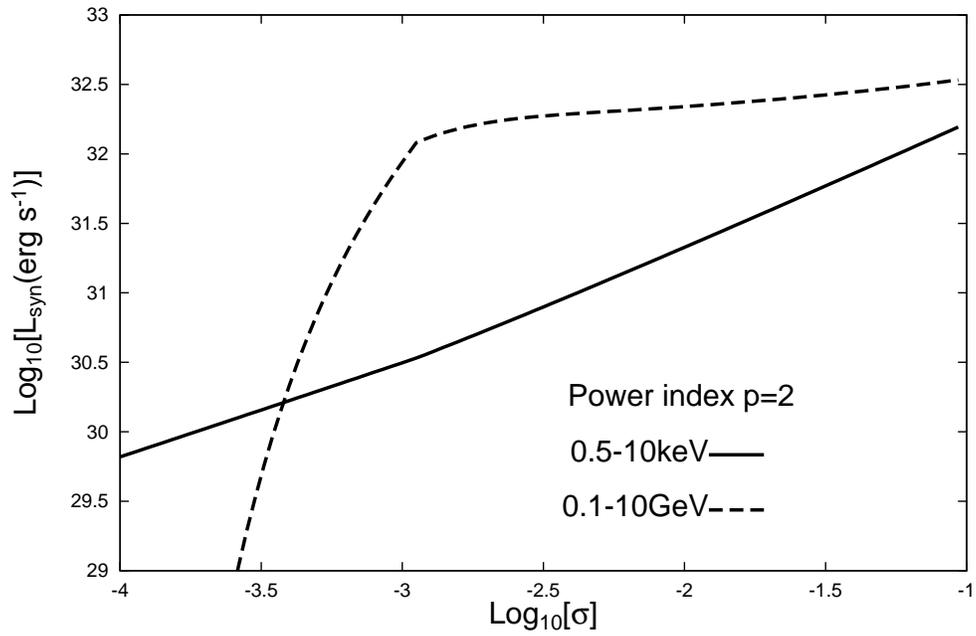}
\caption{The luminosity vs. the magnetization parameter for the 
intra-binary shock model. The solid and dashed lines are the predicted luminosity 
for the 0.5-10~keV energy band and 0.1-10~GeV energy bands, respectively. 
 The results are for the parameters of PSR~B1957+20 and for 
a power law index corresponding to $p\sim2$. }
\label{sigma}
\end{figure}

Figure~\ref{sigma} summarizes the dependence of 
the predicted luminosity of the shock emission in 
  X-ray (solid line) and $\gamma$-ray (dashed line)  bands 
on the magnetization parameter. Here, the results are for 
PSR B1957+20 and for the  power law index of $p\sim2$. 
 For the X-ray bands, we can see 
 in Figure~\ref{sigma} that the luminosity (solid line) 
 decreases with decreasing magnetization parameter. 
This results from the fact that the X-rays are emitted by 
the particles in the slow cooling regime and because  
the magnetic field decreases with decreasing $\sigma$.
In  the 0.1-10~GeV band, the luminosity (dashed-line) is not sensitive to 
 the magnetization parameter if  $\sigma> 10^{-3}$. 
For  $\sigma> 10^{-3}$, the $\gamma$-ray photons are emitted by 
both particles in the slow and fast cooling regimes.  
As the magnetization parameter decreases, the magnetic field 
decreases, while more particles remain in the slow cooling regime. 
Since the former and latter effects tend to decrease and to 
increases the emissivity, respectively, 
two effects compensate  each other. Below   $\sigma\sim 10^{-3}$, 
we can see in Figure~\ref{sigma} that the  $\gamma$-ray
 luminosity quickly  decreases with decreasing magnetization parameter. 
This is because the maximum Lorentz factor
 for $\sigma <10^{-3}$ is limited 
by $\Gamma_g$, at which the gyroradius is equal to size of the system, 
and because  the spectral cut-off of the 
synchrotron radiation, 
$E_{max}\sim 35~L_{34,sd}^{3/2}(\sigma/0.001)^{3/2}r_{s,11}^{-1}$~MeV,
  appears below 100~MeV.

For PSR~B1957+20, 
the non-pulsed X-ray luminosity is observed with 
a luminosity level  of  $L_{X}(0.3-10\mathrm{keV})\sim 2.2\times 
10^{31}~\mathrm{erg~s^{-1}}$ (Huang \& Becker 2007), which can be explained by 
 the intra-binary shock model if the magnetization parameter 
$\sigma> 0.01$.   Guillemot  et al. (2011) find that  
 the X-ray emissions from PSR~B1957+20 are composed of the 
pulsed and non-pulsed components, although  detail spectral 
properties  of the pulsed component is still unknown.  
 Guillemot et al. (2011) also  report the detection of the 
pulsed $\gamma$-ray radiation from  PSR~B1957+20 using  the $Fermi$ data, 
 whose observed flux  level implies the luminosity  of order of 
$L_{\gamma}\sim 1 \times 10^{34}(d/2~\mathrm{kpc})^2
(\Delta\Omega_{\gamma}/4\pi)~\mathrm{erg~s^{-1}}$ (also see Kerr et al. 2010; 
Ray \& Saz Parkinson 2011).  As Table~3 indicates, the predicted luminosity 
level  of the outer gap model (ninth column in Table~3) is  found to be 
 consistent with the $Fermi$ observations with a typical solid angle  
$\Delta\Omega_{\gamma}\sim 3$. 

For other systems, 
 Table~3 indicates that the magnitude of the the predicted luminosity  by 
the intra-binary shock model lies below the results of the  $Fermi$ 
observations,  unless the power law index $p\sim 1.5$. 
For the $\gamma$-ray luminosity of PSR~J1810+17, 
the predictions of  both the outer gap and intra-binary shock models 
 lie below the observation,
 implying that the true spin down power may be larger than 
that assumed using  the relation $\mu_{26}=P^{7/6}_{-3}/3$. 

Due to the uncertainties of  the  magnetization parameter $\sigma$ and the 
power law index $p$, it is  difficult to discriminate between 
the intra-binary shock model and the outer gap model for the unresolved  X-ray 
 emission from the ``black widow'' pulsar,  unless the pulsed period 
is detected in the data.  In addition to the pulsation search,  
future observations may be able to discriminate between models. For example, 
 a  measurement of the spectral shape 
in soft/hard X-ray bands and  $\gamma$-ray bands 
  can discriminate the emission models. 
 The synchrotron spectrum from the intra-binary shock will 
 extend from $E_{min}\sim 200$~eV  to $E_{max}\sim 200$~MeV with a break 
at $E_c\sim 1.6L_{sd,34}^{-3/2}\sigma_{0.1}^{-3/2}r_{s,11}$~MeV, 
  implying the observed spectrum 
can be fit  by a single power law function in the soft/hard X-ray 
bands and by a large photon index above  $\sim100$~MeV. 
For the  outer gap model, on the other hand, 
the  spectral break will appear at $\sim$keV  corresponding to the synchrotron 
radiation  from the secondary pairs,    
as was discussed in section~\ref{nonx},  and at $\sim$GeV corresponding to  the 
curvature radiation in the outer gap. 
 The  observed  soft X-ray spectrum with a index $\alpha\sim 2$,  such 
as the  X-ray emission of PSR~B1957+20 (Huang \& Becker 2007),  
may support the intra-binary shock, although the possibility that
the spectrum with a cut-off energy  $\sim$ 
keV predicted by the outer gap  model can not be excluded.

 Finally, we would like to remark  that (i) many new 
 Black Widow systems  will be associated with  the $Fermi$  un-identified  sources
and (ii) those systems will  exhibit  an eclipse of the radio emission. 
 First, because of the radio eclipse, it is likely that many Black Widow systems  have been missed  
 by the previous radio surveys with the shorter observations. 
On the other hand,  the MSPs in the Black Widow systems are 
younger and have higher spin-down power. 
This indicates that the MSPs in the Black Window 
systems have larger $\gamma$-ray luminosity, 
 $L_{\gamma}\sim f^3L_{sd}$ [equation~(\ref{lgamma})], than ordinary 
MSPs.  Accumulating data of the $Fermi$ observation will 
enable us  to detect  the $\gamma$-ray emissions from the Black Widow systems. 
In particular,  the population studies (e.g. Kaaret \& Philip 1996; Faucher-Gigu$\grave{\mathrm{e}}$re \& Loeb 2010; Takata et al. 2011a,b,c) 
have  pointed out that unidentified MSPs will be associated
 with the $\gamma$-ray sources located at higher Galactic latitude; for example, Takata et al. (2011a,b,c) argued statistically that 
the distribution of the high galactic latitude of the MSPs of 
the $Fermi$ un-identified sources that manifest 
the spectral properties similar to the pulsars can be explained by the 
distribution of the MSPs.   Second,  because the $\gamma$-ray emission from 
the outer gap is greater in the direction perpendicular to the spin axis, 
 $Fermi$ is more likely to discover a greater number of MSPs 
with the Earth viewing angle $\sim 90^{\circ}$ measured 
from the rotation axis (Takata, et al. 2010b; Takata, et al.2011c).
 If  the  angular momentum transferred from the accreting 
 matter to the neutron star  in the accreting 
stage produces the pulsar's spin axis  perpendicular to the orbital plane,
  the $\gamma$-ray emissions from MSPs in Black Widow will be
 greater in the orbital plane. Hence, $Fermi$ will find the 
Black Widow systems with  the Earth viewing angle described by   
 edge-on rather than by face-on with respect 
 to the orbital plane. In such a case,  
a greater number of the  $Fermi$ Black Window systems will reveal 
eclipses of the radio emissions by the matter ejected 
from the companion star.

\section{Summary and conclusion}
\label{summary}
With the recent accumulation of evidence for non-thermal X-ray and $\gamma$-ray emissions 
from different evolutionary stages of MSPs, we have investigated 
 the high-energy emission processes of  isolated rotation powered  
MSPs and those in binary systems.  To understand their observational 
properties, the high-energy emission associated with 
the outer gap accelerator and  with the intra-binary shock in the binary 
system has been investigated. 

For the $\gamma$-ray emitting MSPs,  the polar cap region is heated 
by incoming particles accelerated in the outer gap.  These particles emit $\sim100$MeV photons 
 near the stellar surface which irradiate the polar cap region, 
eventually impacting on the stellar surface.  The former and latter heating processes are 
identified with the rim component characterized by $(T_r,~r_r)\sim (7\times 10^5\mathrm{K},
~1\mathrm{km})$ and the core component  $(T_c,~r_c)\sim (2\times 10^6\mathrm{K},~0.1\mathrm{km})$, 
respectively.  For the outer gap model, the emission properties are controlled by either the 
photon-photon pair-creation process between X-rays from the heated polar cap and 
the $\gamma$-rays or the magnetic pair-creation near the stellar surface.  It has been found, 
based on the statistical grounds, that the outer gap controlled by the magnetic pair-creation 
process near the stellar surface (Takata et al. 2010b) is preferable in explaining the possible 
correlations in $L_{\gamma}$ vs. $L_{sd}$ (or $\tau)$ and $L_{n, X}$ vs. $L_{sd}$. 

For the AMP in the quiescent state of LMXBs, the observed modulation of the optical 
emissions and/or  the non-thermal X-ray emission suggests the presence of   
rotation powered activities during this state. The thermal 
X-ray emission at the neutron star surface resulting from deep crustal heating 
can control the gap and, hence, the $\gamma$-ray emission properties.  We find 
that if the optical modulation originates from the irradiation of $\gamma$-rays 
from the outer gap, the observed amplitude can constrain the  
 NS model. For example,  the level of the inferred irradiation luminosity 
$L_{irr}\sim 10^{34}~\mathrm{erg~s^{-1}}$ of SAX~J1808.4-3658 suggests 
that the outer gap with a high-mass NS model is preferable.  
As argued by Takata et al. (2010a), the presence of the outer gap emission 
would be responsible for the transition of 
the system from the LMXB phase to the rotation powered MSP phase. 

Finally, we have discussed the high-energy emission from an intra-binary shock 
in the black widow systems,  which are  frequently  found from radio searches of 
$Fermi$ unidentified sources.  Within the context of the simple one-zone model, 
we calculate the synchrotron emission from the accelerated particles at 
the shock.  For PSR~B1957+20, the observed  non-pulsed X-ray 
emission ($L_{X}\sim 2.2\times 10^{31}\mathrm{erg~s^{-1}}$) can 
be explained by the intra-binary shock model if the magnetization 
parameter is $\sigma>0.01$.  In addition, it is found that the observed luminosity  
of the  pulsed $\gamma$-ray emission from PSR~B1957+20 can be explained 
by the outer gap model. On the other hand, the 0.1-10~GeV  emissions from the 
intra-binary  shock is several orders of magnitude smaller than that from the outer gap 
emission, unless the magnetization parameter and the power law index 
of the accelerated particles are $\sigma\ge 10^{-3}$ and  
$p\sim 1.5$, respectively.  For the other black widow systems detected 
from $Fermi$ unidentified sources, the predicted 
luminosity from the intra-binary shock model is consistent with 
 the $Fermi$ observations  only 
if the power law index $p\sim 1.5$.  In addition to the pulse search, 
the origin of the high-energy emissions from black widow systems 
will be constrained by a measurement of the spectral shape   
by new studies in the soft/hard X-ray bands, for example, by the Astro-H satellite, (Takahashi 
et al. 2010) ) and/or by $Fermi$.

\acknowledgements 
 We thank   A.H.Kong, C.Y.Hui,  P.H.T.Tam, R.H.H.Huang, 
 Lupin-C.C.Lin, M.Ruderman, and  S.Shibata for the useful discussions. 
We express our appreciation to an anonymous referee for  useful  comments.
J.T. and K.S.C. are supported by a GRF grant of Hong Kong Government  under 
HKU700911P and R.E.T This are  supported in part by 
the Theoretical Institute for Advanced 
Research in Astrophysics (TIARA) operated under the Academia Sinica Institute of Astronomy \& 
Astrophysics in Taipei, Taiwan.


\newpage

\begin{landscape}
\begin{table}
\begin{tabular}{ccccccccccc}
\raisebox{-1.8ex}[0pt][0pt]{Pulsar name} & $P$ & $\mu$ & $\tau_c$ &$L_{sd}$  & $d$ & $T_r
$ & $R_r$ & $T_c$ & $R_c$ & $L_{\gamma}$ \\
 & (ms) & ($10^{26}~\mathrm{G~cm^3}$) & ($10^9$~yrs)& $(10^{34} \mathrm{erg~s^{-1}}$) & (kpc) & ($10^6$~K) & (km
) & ($10^6$~K) & (km) & $(10^{32}~\mathrm{erg s^{-1}})$ \\
\hline\hline
J0030+0451 & 4.9 & 2.3& 7& 0.35 & 0.3& 0.7 &2 & 1.5 & $0.4^{(a)}$ & $5.7\pm 3.5$ \\
J0034-0534 & 1.9 & 0.7 & 11& 2.9 & 0.53 &  &  & 2.2 & $0.05^{(b)}$ & $4.7\pm 3.8$ \\
J0218+4232 & 2.3 & 4.1 & 0.5& 24.6 & 2.7 &  &  & 2.9 & $0.37^{(c)}$ & 270-690 \\
J0437-4715 & 5.8 & 3 & 6&0.29 & 0.156 & 0.52 & 2.9 & 1.4 & $0.39^{(b)}$
 & $0.54\pm 0.08$ \\
J0613-0200 & 3.1 & 1.8 &4& 1.3 & 0.48 &  &  &  &  & $8.9^{+7.1}_{-4.2}$ \\
J0751+1807 & 3.5 & 1.5 &8& 0.7 & 0.62 &  &  & 3.7 & $0.04^{(d)}$ & $4.7^{+10}_{-3.5}$ \\
J1614-2230 & 3.2 & 1.2 &11& 0.5 & 1.3 &  &  & 2.7 & $0.12^{(e)}$ & $53\pm 34$ \\
J1744-1134 & 4.1 & 1.8 &8& 0.4 & 0.47 &  &  & 3.2 & $0.03^{(e)}$ & $4.3\pm 1.3$ \\
B1937+21 & 1.6 & 4.1 &0.2 & 110 & 7.7 &  &  &  &  & 560-2500 \\
B1957+20 & 1.6 & 1.7 &1& 11 & 2.5 &  &  &  &  & $\sim$100 \\
J2124-3358 & 4.9 & 2.4 &6& 0.4 & 0.25 & 0.5 & 1.8 & 2.2 & $0.1^{(b)}$ & $2.1^{+4.2}_{-1.4}$ \\
\hline
\end{tabular}
\caption{Observational properties of the millisecond pulsars detected by 
$Fermi$-LAT. $P$; pulsar spin period, $\mu$; magnetic moment of 
the neutron star, $\tau_c$; spin down age,  $L_{sd}$; pulsar spin down power, $d$; distance from the Earth, $T_r$; surface temperature 
of the rim component, $R_r$; effective radius of the rim component,  $T_c$; surface temperature of the core component, $R_c$; effective radius of the core component and $L_{\gamma}$; $\gamma$-ray luminosity. 
For the $\gamma$-ray luminosity, the data were taken from Abdo et al. (2009b). 
For the blanks in  the columns $T_r\sim R_c$, no available data was found; 
(a) Bogdanov \& Grindlay (2009), (b)Zavlin (2006), (c)
Webb,  Olive \&  Barret (2004), (d)Webb et al. (2004), and 
(e)Marelli, De Luca \&  Caraveo (2011).}
\end{table}
\end{landscape}
\newpage 

\begin{landscape}
\begin{table}
\begin{tabular}{cccc|cccc|ccccl}
\raisebox{-1.8ex}[0pt][0pt]{Pulsar name} & $T^{ob}_r$ & $T^{ob}_{c}$  & $L^{ob}_{\gamma}$ & \raisebox{-1.8ex}[0pt][0pt]{$f_p$} & $T^p_{r}$ & $T_{c}^{p}$  & $L_{\gamma}^{p}$ & \raisebox{-1.8ex}[0pt][0pt]{$f_m$} & $T^m_{r}$ & $T_{c}^{m}$  & $L_{\gamma}^{m}$ &  \\
 & \multicolumn{2}{c}{($10^{6}$~K)} & \multicolumn{1}{c|}{($10^{32}~\mathrm{erg s^{-1}}$)} & \multicolumn{1}{c}{} & \multicolumn{2}{c}{($10^{6}$~K)} & \multicolumn{1}{c|}{($10^{32}~\mathrm{erg s^{-1}}$)} & \multicolumn{1}{c}{} & \multicolumn{2}{c}{($10^{6}$~K)} & \multicolumn{1}{c}{($10^{32}~\mathrm{erg s^{-1}}$)} &  \\
\cline{1-12}
\hline\hline
J0030+0451 & 0.7 & 1.5 & $5.7\pm3.5$ & 0.51 & 0.7 & 1.2 & 4.6 & 0.55 & 0.72 & 1.3 & 5.9 &  \\
J0034-0534 &  & 2.2 & $4.7\pm3.8$ & 0.37 & 0.55 & 3.8 & 6.5 & 0.34 & 0.56 & 3.6 & 5.8 &  \\
J0218+4232 &  & 2.9 & 270-690 & 0.16 & 0.66 & 1.6 & 10 & 0.38 & 0.82 & 2.0 & 124 &  \\
J0437-4715 & 0.52 & 1.4 & $0.54\pm0.08$ & 0.60 & 0.6 & 1.2 & 6.5 & 0.60 & 0.6 & 1.2 & 6.6 &  \\
J0613-0200 &  &  & $8.9^{+7.1}_{-4.2}$ & 0.38 & 0.59 & 2.7 & 7.1 & 0.44 & 0.62 & 2.8 & 11.3 &  \\
J0751+1807 &  & 3.7 & $4.7^{+10}_{-3.5}$ & 0.48 & 0.5 & 4.0 & 6.5 & 0.47 & 0.57 & 3.9 & 5.8 &  \\
J1614-2230 &  & 2.7 & $53\pm34$ & 0.49 & 0.56 & 2.3 & 6.3 & 0.44 & 0.56 & 2.2 & 4.7 &  \\
J1744-1134 &  & 3.2 & $4.3\pm1.3$ & 0.53 & 0.58 & 4.6 & 6.5 & 0.50 & 0.58 & 4.4 & 5.7 &  \\
B1937+21 &  &  & 560-2500 & 0.10 & 0.7 & 3.2 & 10 & 0.31 & 0.91 & 4.4 & 330 &  \\
B1957+20 &  &  & $\sim 100$ & 0.17 & 0.62 & 2.9 & 5.4 & 0.32 & 0.73 & 3.5 & 36 &  \\
J2124-3358 & 0.5 & 2.2 & $2.1^{+4.2}_{-1.4}$ & 0.48 & 0.73 & 2.4 & 4.3 & 0.55 & 0.76 & 2.5 & 6.4 &  \\
\hline
\end{tabular}
\caption{The  X-ray and $\gamma$-ray emissions from rotation powered MSPs. 
$T^i_r$ and  $T^i_c$ represent the temperatures of the rim and core 
components of the thermal X-ray emissions, respectively, 
and $L^i_{\gamma}$ is the $\gamma$-ray luminosity. The superscript represents
 the observations ($i=ob$), the predictions of the outer gap model 
controlled by the photon-photon pair-creation process ($i=p$)
 and by the magnetic pair-creation process ($i=m$), respectively. 
 For the outer gap model 
controlled by the  photon-photon pair-creation process ($i=p$), 
the gap fraction ($f_p$), the temperatures ($T_{r}^{p},~T_c^p$) and the
 luminosity ($L_{\gamma}^p$) are calculated from equations~(\ref{fp}),  
(\ref{ptr}), (\ref{ptc}), and (\ref{plgamma}), respectively, where 
 we used (1) the observed effective 
radius if they were reported
or (2) the typical value $R_r=3\times 10^5$~cm and 
$R_c=10^4$~cm if there were no available observation results. 
In addition, we used the curvature radii corresponding to
 $s_*=0.5$, and $\zeta_6=1$. For the outer gap model controlled by the magnetic pair-process $i=m$,  the gap fraction  $f_m$, the temperatures ($T_{r}^{m},~T_c^m$) and the luminosity ($L_{\gamma}^m$) are calculated from 
equations~(\ref{fm}), (\ref{mtr}), (\ref{mtc}) and (\ref{mlgamma}), 
respectively, where we used $K_1=1$. }

\end{table}
\end{landscape}

\begin{landscape}
\begin{table}
\begin{tabular}{cccccccccc}
\raisebox{-1.8ex}[0pt][0pt]{Pulsar name$^a$} & P & $L_{sd}$ & $L^{ob}_{\gamma}$ & $P_b$ & $a_o$ &$L^s_{\gamma}$ & $L^s_{X}$ & $L^m_{\gamma}$ & $L^m_{X}$ \\
 & (ms) & $10^{34}\mathrm{erg~s^{-1}}$&$10^{32}~\mathrm{erg~s^{-1}}$ & hr & $R_{\odot}$ &$ 10^{32}~\mathrm{erg~s^{-1}}$ &$10^{31}~\mathrm{erg~s^{-1}}$ & $10^{32}~\mathrm{erg~s^{-1}}$ & $10^{31}~\mathrm{erg~s^{-1}}$ \\
\hline\hline
J0023+09(F) & 3.05 & $0.66^{\dagger}$ &8 & 3.3 & 1.3 &0.5,~0.02& 0.05,~0.23 & 5.5 & 0.04 \\
J0610-21 & 3.86 & 0.23& & 6.9 & 2.1&0.1,~0.004  & 0.004,~0.02& 2.7&  0.01 \\
J1731-1847 & 2.3 & $1.1^{\dagger}$& & 7.5 & 2.2 & 1,~0.003
& 0.09,~0.4 &6 & 0.07 \\
J1745+30(F) & 2.65 & 1.3& $19^{\ddagger}$ & 17.5 & 3.8 & 1,~0.003 & 0.09,~0.35 
 &8.8 & 0.1 \\
J1810+17(F) & 1.66 & $1.8^{\dagger}$& 120 & 3.6 & 1.3 &2,~0.008& 0.35,~1.5 & 6 & 0.1 \\
B1957+20(F) & 1.61 & 11 &100 & 9.2 & 2.5 &20,~0.05& 9.3,~26& 36 & 1 \\
J2051-0827 & 4.51 & 0.53& & 2.4 & 1&3.9,~0.02 & 0.03,~0.2 & 8 & 0.05 \\
J2214+30(F) & 3.12 & 1.9&88 & 10.0 & 2.6 & 2,~0.006& 0.2,~1&16 & 0.2 \\
J2241-52(F) & 2.19 & 3.3 & 10& 3.4 & 1.3 &4,~0.016& 1.2,~4.4 & 17 & 0.3 \\
J2256-1024(F) & 2.29 & 5.2 &4.3& 5.1 & 1.7  &7,~0.025& 2.6,~8.6& 28 & 0.5 \\
\hline
\end{tabular}
\caption{
The predicted X-ray and $\gamma$-ray emissions 
from ``black widow'' MSPs. The MSPs associating with $Fermi$ unidentified 
sources are denoted as (F) in the first column.  The second ($P$) and third ($L_{sd}$) 
 columns are the spin period and the spin down power of the pulsar, respectively; 
$\dagger$ we assumed 
$\mu_{26}=P^{7/6}_{-3}/3$ (dashed-line in Figure~\ref{pb}) where 
the dipole moment of the neutron star has not been measured.
The fourth column represents  the expected $\gamma$-ray luminosity, $L^{ob}_{\gamma}=4\pi d^2F_{>0.1}$, where 
 $F_{>0.1}$ is the observed flux above 100~MeV,  
 and $d$ is the distance to the black widow system;  $\ddagger$ we assumed 
$d=1$~kpc because of no available measurement. 
The fifth ($P_o$) is the orbital period, and the sixth column ($a_o$) is 
 inferred  separation (in units of the solar radius) of the two components
 if the Earth viewing angle is edge-on to the orbital plane. 
The seventh $L^s_{X}$ and eighth $L^s_{\gamma}$ columns represent 
the  luminosity  in the 0.5-10~keV and in 0.1-10~GeV energy bands, 
respectively, predicted by  the intra-binary shock model [Eq.~\ref{inx-ray}], 
where we assumed the shock distance $r_s=a_0$, $\eta=0.03$  and  $\sigma=0.1$.
 In addition,  the first and the second values shows the results for 
the power law index $p=1.5$ and $p=3$, respectively.
 The ninth   ($L^m_{\gamma}$) and tenth ($L^m_{X}$) column  show 
 the predicted $\gamma$-ray~[Eq.(\ref{mlgamma})]  
and X-ray~[Eq.(\ref{nxm})] radiations from  the outer gap model 
 controlled by the magnetic pair-creation model, respectively, where 
we assumed $K_1=1$, $R_{r,5}=3$, and $\zeta_6=1$. 
}
\end{table}
\end{landscape}


\begin{thebibliography}{}
\bibitem[\protect\citeauthoryear{Abdo}{2011}]{abo11}
Abdo A.A. et al. 2011, ApJL, 736, 11
\bibitem[\protect\citeauthoryear{Abdo}{2010}]{ab010a}
Abdo A.A. et al. 2010a, ApJS,  187, 460
\bibitem[\protect\citeauthoryear{Abdo}{2010}]{ab010b}
Abdo A.A. et al. 2010b, ApJS,  188, 405
\bibitem[\protect\citeauthoryear{Abdo}{2010}]{ab010c}
Abdo A.A. et al. 2010c, ApJ,  708, 1426
\bibitem[\protect\citeauthoryear{Abdo}{2010}]{ab010d}
Abdo A.A. et al. 2010d, ApJ, 712, 957
\bibitem[\protect\citeauthoryear{Abdo}{2009}]{ab09a}
Abdo A.A. et al. 2009a, Sci., 325, 840
\bibitem[\protect\citeauthoryear{Abdo}{2009}]{ab09b}
Abdo A.A. et al. 2009b, Sci., 325, 848
\bibitem[\protect\citeauthoryear{Abdo}{2009}]{ab09c}
Abdo A.A. et al. 2009c, ApJ, 706, 1331
\bibitem[\protect\citeauthoryear{Aliu}{2008}]{al08}
Aliu, E. et al. 2008, Sci, 322, 1221
\bibitem[\protect\citeauthoryear{Alpar}{1982}]{al82} Alpar, M.A., Cheng, A.F., Ruderman, M.A. \& Shaham, J., 1982, 
Natur, 300, 728
\bibitem[\protect\citeauthoryear{Aharonion}{2009}]{ah09}
Aharonian, F., et al., 2009, A\&A, 507, 389
\bibitem[\protect\citeauthoryear{Arons}{1983}]{ar83}
Arons J., 1983 ApJ, 266, 215
\bibitem[\protect\citeauthoryear{Arons}{1994}]{ar94}
Arons J., \& Tavani, M., 1994, 403, 249
\bibitem[\protect\citeauthoryear{Baring}{2004}]{Ba04}
Baring, M.G., 2004, NuPhS., 136, 198
\bibitem[\protect\citeauthoryear{Bogdanov}{2009}]{bo09}Bogdanov, S. \&  Grindlay, J.E., 2009, ApJ, 703, 1557 
\bibitem[\protect\citeauthoryear{Brown}{1990}]{Br90}Brown, E.F., Bildsten, L. \&  Rutledge, R.E. 1998, ApJL, 504, 95
\bibitem[\protect\citeauthoryear{Burderi}{2003}]{Bu03} Burderi, L., Di Salvo, T., D'Antona, F., Robba, N.R., \& 
Testa, V. 2003, \aap, 404, L43
\bibitem[\protect\citeauthoryear{Campana}{1990}]{Ca90} Campana, S., Colpi, M, Mereghetti, S., Stella, L., \& Tavani, M.
 1998,  A\&ARv, 8, 279
\bibitem[\protect\citeauthoryear{Chakrabarty}{1998}]{Ch98} Chakrabarty, D., Morgan, E.H., 1998,  Natur, 394, 346
\bibitem[\protect\citeauthoryear{Cheng}{1998}]{ch98} Chen, K., Ruderman, M., \& Zhu, T. 1998, ApJ, 493, 397
\bibitem[\protect\citeauthoryear{Cheng}{2006}]{ch06} Cheng, K. S., Taam, R.E., \& Wang, W., 2006, ApJ, 641, 427
\bibitem[\protect\citeauthoryear{Cheng}{1986a}]{ch86a} Cheng, K. S., Ho, C., \& Ruderman, M. 1986a, \apj, 300, 500
\bibitem[\protect\citeauthoryear{Cheng}{1986b}]{ch86b} Cheng, K. S., Ho, C., \& Ruderman, M. 1986b, \apj, 300, 522
\bibitem[\protect\citeauthoryear{Cheng}{1999}]{ch99} Cheng, K. S., \& Zhang, L., 1999, \apj, 515 337
\bibitem[\protect\citeauthoryear{Cheng}{2000}]{ch00} Cheng, K. S., Ruderman, M., \& Zhang, L., 2000, APJ,  537, 964
\bibitem{co01}Colpi, M., Geppert, U., Page, D., \&  Possenti, A. 2001, 
ApJL, 548, 175
\bibitem{ca09} D'Avanzo, P., Campana, S., Casares, J., Covino, S., Israel, G.L.,
 \& Stella, L. 2009, \aap, 508, 297
\bibitem[\protect\citeauthoryear{Daugherty}{1996}]{da96}
Daugherty J.K. \&  Harding, A.K. 1996, ApJ, 458, 278
\bibitem[\protect\citeauthoryear{Daugherty}{1982}]{da82}
Daugherty J.K., Harding, A.K. 1982, ApJ, 252, 337
\bibitem[\protect\citeauthoryear{Deloye}{2008}]{De08} Deloye, C. J., Heinke, C. O., Taam, R. E., 
\& Jonker, P. G. 2008, \mnras, 391, 1619
\bibitem[\protect\citeauthoryear{Faucher}{2010}]{fa10}
 Faucher-Gigu$\grave{\mathrm{e}}$re, C.-A., \&   Loeb, A. 2010, J. Cosmol. 
 Astropart. Phys., JCAP, 1, 5
\bibitem{fr02} Frank, J., King, A., \& Raine, D. 2002, Accretion Power in Astrophysics (Cambridge: Cambridge Univ. Press)
\bibitem{gu83}Gudmundsson, E.H., Pethick, C.J. \&  Epstein, R.I. 1983, ApJ, 272, 286
\bibitem{gu11}Guillemot,L., et al. 2011, arXiv1110.1271
\bibitem{ha03}Haensel, P. \&  Zdunik, J.L. 2003, A\&AL, 404, 33
\bibitem{ha90}Haensel, P. \&  Zdunik, J.L. 1990, A\&A ,227, 431
\bibitem{ha93}Halpern, J.P. \& Ruderman, M. 1993, ApJ, 415, 286
\bibitem[\protect\citeauthoryear{Harding}{2011}]{ha11}
Harding, A. K. \&  Muslimov, A.G. 2011, ApJL, 726, 10
\bibitem[\protect\citeauthoryear{Harding}{2008}]{ha08}
Harding, A.K., Stern, J.V.; Dyks, J., Frackowiak, M. 2008, ApJ, 680, 1378
\bibitem[\protect\citeauthoryear{Harding}{2005}]{ha05}
Harding, A. K., Usov, V.V., \&  Muslimov, A.G. 2005, ApJ, 622, 531
\bibitem{he09}Heinke, C.O.,  Cohn, H.N. \& Lugger, P.M. 2009, ApJ, 692, 584
\bibitem[\protect\citeauthoryear{Hirotani}{2008}]{hi08}
Hirotani K. 2008, ApJL, 688, 25
\bibitem{hu07}Huang, H.H. \& Becker, W.,  2007, A\&AL, 463, 5
\bibitem[\protect\citeauthoryear{ka}{1996}]{ka96}
Kaaret, P., \&  Cottam, J., 1996, ApJL, 462, 35
\bibitem{jo08} Jonker, P. G., Torres, M. A. P., \& Steeghs, D. 2008, 
\apj, 680, 615
\bibitem{ka11}Kalapotharakos, C., Kazanas, D., Harding, A., \& 
 Contopoulos, I., 2011, arXiv1108.2138
\bibitem{ka05} Kargaltsev, O.Y., Pavlov, G.G., Zavlin, V.E. \& Romani, R.W.,  
2005, ApJ, 625, 307
\bibitem[\protect\citeauthoryear{keith}{2011}]{ke11}
Keith, M.J. et al. 2011, MNRAS, 414, 1292
\bibitem{ke10} Kerr, M., Fermi LAT Collaboration, \& Pulsar Timing Consortium. 
2010, in  AAS/High Energy Astrophysics Division Meeting, Vol. 11, 23.03
\bibitem{ko11}Kong, S.W., Yu, Y.W., Huang, Y.F. \& 
 Cheng, K. S., 2011, MNRAS, 416, 1067
\bibitem{la99} Lang K.R. 1999, Astrophysical Formulae (New York: Springer)
\bibitem{ly01}Lyubarsky, Y., \&  Kirk, J.G., 
 2001, ApJL, 547, 437
\bibitem{ma11}Marelli, M., De Luca, A., \&  Caraveo, P.A., 2011, ApJ, 733, 82
\bibitem[\protect\citeauthoryear{Muslimov}{2004}]{mu04}
Muslimov, A.G. \& Harding, A.K. ApJ, 617, 471
\bibitem{ra11}Ray, P. \& Saz Parkinson, P.M., 2011, in Rea N., Torres D. F., eds, High-Energy Emission from Pulsars and Their Systems. Springer, Berlin, p. 37
\bibitem{ro10}Roberts, et al., 2011, to appear in AIP Conference Proceedings of Pulsar Conference 2010 "Radio Pulsars: a key to unlock the secrets of the Universe", Sardinia, October 2010 (arXiv:1103.0819) 
\bibitem[\protect\citeauthoryear{Ruderman}{1991}]{ru91}
Ruderman M. 1991, ApJ, 366, 261	
\bibitem[\protect\citeauthoryear{Ruderman}{1975}]{ru75}
Ruderman M.A.,  Sutherland P.G. 1975, ApJ, 196, 51
\bibitem{ro09} Romanova, M.M., Ustyugova, G.V., Koldoba, A.V., \& Lovelace,
 R.V.E. 2009, \mnras, 399, 1802
\bibitem[\protect\citeauthoryear{Parkinson}{2010}]{pa10}
Saz~Parkinson, P.M. et al. 2010, 725, 571
\bibitem{sp06}  Spitkovsky A., 2006,648, 51
\bibitem{st03}Stappers, B.W., Gaensler, B.M., Kaspi, V.M., van der Klis, M., \& 
 Lewin, W.H.G., 2003, Sci, 299, 1372
\bibitem[Takahashi(2010)]{Ta10}
Takahashi, et al.. 2010. Proceedings of the SPIE, 7732, 77320Z-77320Z-18 
\bibitem[\protect\citeauthoryear{Takata}{2011a}]{ta11a}
Takata, J., Wang, Y.,  Cheng, K.S. 2011a, ApJ, 726, 44
\bibitem[\protect\citeauthoryear{Takata}{2011b}]{ta11b}
Takata, J., Wang, Y.,  Cheng, K.S. 2011b, MNRAS, 414,2173
\bibitem[\protect\citeauthoryear{Takata}{2011b}]{ta11c}
Takata, J., Wang, Y.,  Cheng, K.S. 2011c, MNRAS, 415, 1827
\bibitem[\protect\citeauthoryear{Takata}{2010}]{ta10a}
Takata, J., Cheng, K.S.,  Taam, R.E. 2010a, ApJL, 723, 68
\bibitem[\protect\citeauthoryear{Takata}{2010}]{ta10b}
Takata, J., Wang, Y.,  Cheng, K.S. 2010b, ApJ, 715, 1318
\bibitem{ta09}Takata, J., \& Taam, R.E., 2009, ApJ, 702, 100
\bibitem{ta08}Takata, J., Chang, H.-K. \& Shibata, S, 2008, MNRAS, 386, 748
\bibitem{ta11}
Tam, P.H.T., Huang, R.H.H., Takata, J., Hui, C.Y., Kong, A.K.H. \& Cheng, K.S.,
 2011, ApJL, 736, 10 
\bibitem[\protect\citeauthoryear{Tam}{2010}]{tam10}
Tam, P.H.T. et al. 2010, ApJL, 724, 207
\bibitem{ta97}
Tavani, M, \&  Arons, J., 1997, ApJ, 477, 439
\bibitem{th05} Thorstensen, J. R., \& Armstrong, E. 2005, \aj, 130, 759
\bibitem[\protect\citeauthoryear{Venter}{2009}]{ve09}
Venter C.,  Harding A.K.,  Guillemot L. 2009, ApJ, 707, 800
\bibitem{wa11}Wada, T.,  \& Shibata, S., 2011, MNRAS, in press
\bibitem{wa98}Wang, F. Y.-H., Ruderman, M., Halpern, J.P. \& Zhu, T., 
1998, ApJ, 498, 373
\bibitem[\protect\citeauthoryear{Wang}{2010}]{wa10}
Wang, Y.,  Takata, J. \&  Cheng, K.S. 2010, ApJ, 720, 178
\bibitem{wa09}  Wang, Z., Archibald, A. M., Thorstensen, J. R., Kaspi, V. M., 
Lorimer, D. R.,  Stairs, I., \& Ransom, S. M. 2009, \apj, 703, 2017
\bibitem {we04} Webb, N. A., Olive, J.-F., Barret, D., Kramer, M., Cognard, I., 
 L$\mathrm{\ddot{o}}$hmer, O., 2004, A\&A, 419, 269
\bibitem{we04} Webb, N.A., Olive, J.-F., Barret, D., 2004 A\&A, 417, 181
\bibitem{ya04} Yakovlev, D.G. \&  Pethick, C.J. 2004 ARA\&A, 42, 169
\bibitem{ya03} Yakovlev, D.G., Levenfish, K.P.\&  Haensel, P. 2003, A\&A, 407, 265
\bibitem{za06} Zavlin, V.E. 2006, ApJ, 638, 951
\bibitem{za07} Zavlin, V.E. 2007, Ap\& SS, 308, 297 
\bibitem{zh03} Zhang, L. \& Cheng, K.S. 2003, A\&A, 398, 639
\bibitem{zh97} Zhang, L. \& Cheng, K.S. 1997, \apj, 480, 370
\end{thebibliography}
\end{document}